\RequirePackage{rotating}
\documentclass{mnras}

\usepackage{graphicx,txfonts,times}

\usepackage{amsmath,amssymb}
\usepackage{natbib}
\usepackage{epsfig}
\usepackage{color,float}
\usepackage[usenames,dvipsnames,svgnames,table]{xcolor}
\usepackage{enumerate}
\usepackage{epstopdf}
\usepackage{booktabs}
\DeclareGraphicsExtensions{.pdf,.png,.jpeg,.ps,.eps}
\usepackage{epsfig}
\usepackage{threeparttable}
\usepackage{framed}
\usepackage[T1]{fontenc}
\usepackage{aecompl}
\usepackage{soul}
\usepackage{subcaption}
\usepackage{ltablex}
\usepackage{pdflscape}
\usepackage{xtab, afterpage}
\usepackage{hyperref}

\newcommand{\lclash}{Cluster Lensing And Supernova survey with Hubble}
\newcommand{\sbu}{mag arcsec$^{-2}$}

\newcommand{\hst}{\textit{HST}}
\newcommand{\chandra}{\textit{Chandra}}
\newcommand{\XMM}{\textit{XMM-Newton}}

\newcommand{\ezgal}{\texttt{EZGAL}}

\newcommand{\AD}{\texttt{AstroDrizzle}}
\newcommand{\mfive}{M$_{500,c}$}

\newcommand{\rfive}{r$_{500,c}$}

\newcommand{\blue}{F105W}
\newcommand{\red}{F160W}
\newcommand{\green}{F110W}
\newcommand{\til}{$\thicksim$}

\newcommand{\zform}{z$_{f}$}
\newcommand{\Lstar}{L*}
\newcommand{\Lsun}{L$_{\odot}$}

\newcommand{\Msun}{M$_\odot$}
\newcommand{\Mstar}{M$_*$}
\newcommand{\Mstell}{M$_\bigstar$}

\newcommand{\colorgrad}{d(F110W-F160W)(d $\log$(r))$^{-1}$}
\newcommand{\teneleven}{$\times$10$^{11}$}
\newcommand{\tentwelve}{$\times$10$^{12}$}
\newcommand{\tenthirteen}{$\times$10$^{13}$}
\newcommand{\tenfourteen}{$\times$10$^{14}$}
\newcommand{\tenfifteen}{$\times$10$^{15}$}
\newcommand{\dlogr}{dlog(r[kpc])}
\newcommand{\nab}{$\nabla_{F110W-F160W}$}
\newcommand{\logmfive}{log(M$_{500,c}$ [M$_\odot$])}
\newcommand{\logmstell}{log(M$_\bigstar$/M$_\odot$)}

\def\aj{\rmfamily{AJ}}
\def\apj{\rmfamily{ApJ}}
\def\apjl{\rmfamily{ApJ}}
\def\apjs{\rmfamily{ApJS}}

\def\aap{\rmfamily{A\&A}}

\def\aaps{\rmfamily{A\&AS}}
\def\mnras{\rmfamily{MNRAS}}
\def\pasp{\rmfamily{PASP}}
\def\ao{\rmfamily{Appl.~Opt.}}

\title[Lost but not Forgotten]
{Lost but not Forgotten: Intracluster Light in Galaxy Groups and Clusters }

\author[T. DeMaio et al.]{
Tahlia DeMaio,$^{1}$
Anthony H. Gonzalez,$^{1}$
Ann Zabludoff,$^{2}$
Dennis Zaritsky,$^{2}$
\newauthor
Thomas Connor, $^{3, 4}$
Megan Donahue, $^{4}$
and John S. Mulchaey$^{3}$ 
\\
$^{1}$Department of Astronomy, University of Florida, Gainesville, FL 32611\\
$^{2}$Department of Astronomy, University of Arizona, Steward Observatory, Tucson, AZ  85721\\
$^{3}$The Observatories of the Carnegie Institution for Science, 813 Santa Barbara St, Pasadena, CA 91101\\
$^{4}$Department of Physics and Astronomy, Michigan State University, East Lansing, MI 48824
}

\begin{document}

\maketitle

\begin{abstract}
With \emph{Hubble Space Telescope} imaging, we investigate the progenitor population and formation mechanisms of the intracluster light (ICL) for 23 galaxy groups and clusters ranging from 3\tenthirteen$<$\mfive\ [\Msun]$<$9\tenfourteen\ at 0.29$\leq$z$\leq$0.89.
The colour gradients of the BCG+ICL get bluer with increasing radius out to 53-100 kpc for all but one system, suggesting that violent relaxation after major mergers with the BCG cannot be the dominant source of ICL. 
The average colour gradient for clusters with \mfive$>$1\tenfourteen\ \Msun\ is not statistically different than that of the lower-mass groups.
The BCG+ICL luminosity within 100 kpc increases with total cluster mass more steeply than within 10 kpc, implying a decoupling between in the inner and outer stellar components.
For clusters the BCG+ICL luminosity at r$<$100 kpc (0.08-0.13 \rfive) is 1.2$-$3.5\tentwelve\ \Lsun; for the groups, BCG+ICL luminosities within 100 kpc  (0.17-0.23 \rfive) range between  0.7-1.3\tentwelve\ \Lsun. 
The BCG+ICL stellar mass in the inner 100 kpc increases with total cluster mass as \Mstell$\propto$\mfive$^{0.37\pm0.05}$.
This steep slope implies that the BCG+ICL makes up a higher fraction of the total mass in groups than in clusters and that group environments are more efficient ICL producers within 100 kpc. 
The BCG+ICL luminosities and stellar masses are too large for the ICL stars to come from the dissolution of dwarf galaxies alone, given the observed evolution of the faint end of the cluster galaxy luminosity function, implying instead that the ICL grows from the stripping of more massive galaxies.
Using the colours of cluster members from the CLASH high-mass sample, we place conservative lower limits on the luminosities of galaxies from which the ICL at r$<$100 kpc could originate via stripping.
We find that the ICL has a colour similar to massive, passive cluster galaxies ($>$10$^{11.6}$ \Msun) at 10 kpc, while by 100 kpc this colour is equivalent to that of a 10$^{10.0}$ \Msun\ galaxy.
Additionally, the colour of the BCG+ICL light within 100 kpc is consistent with 75\% of the total BCG+ICL luminosity originating in galaxies with L$>$0.2 \Lstar\ (log(\Mstell\ [\Msun])$>$10.4), assuming conservatively that these galaxies are completely disrupted.
We conclude that the tidal stripping of massive galaxies is the likely source of the intracluster light from 10$-$100 kpc (0.008-0.23 \rfive) for galaxy groups and clusters.
\end{abstract}


\section{Introduction}
Intracluster light (ICL) is the diffuse, low surface brightness component of galaxy groups and clusters.
It is composed of stars that are not bound to an individual galaxy but are instead associated with the cluster potential.
Any star that becomes unbound from its parent galaxy in the cluster remains in the ICL, making the ICL a fossil record of all past interactions.
The colour, metallicity, spatial distribution, and surface brightness of the ICL reflect the properties of galaxies in which the intracluster stars originated, effectively encoding the formation history of the cluster.
Not only is the formation of the ICL closely linked to the process of cluster assembly \citep{Rudick2006}, it also offers a way to constrain how galaxies evolve and interact in the dense environments of galaxy groups and clusters. 
Each formation mechanism affects the distribution of intracluster stellar populations in different ways.
We can use observations of the colours of galaxies and the colour distribution of the ICL to discern which mechanisms play the largest roles in the build-up of the ICL. 
The three main channels for ICL build-up and their effects on the ICL colour gradient are:

\begin{enumerate}[(1)]
 \item Complete dwarf disruption: Low mass, and thus low-metallicity, \citep{Zaritsky1994a, Skillman1996a} dwarfs can be completely shredded by cluster tidal forces.
The depth within the cluster potential at which each dwarf is shredded depends on the mass of the dwarf, forming a colour gradient as bluer, lower-mass dwarfs are disrupted at larger cluster radii compared to more massive, more metal-rich galaxies \citep{Rudick2010, Melnick2012, Conroy2007}.

\item Partial tidal stripping: Tidal interactions play a significant role in galactic evolution. 
Tidal interactions liberate stars from these galaxies and deposit them in the ICL. 
Because galaxies have internal colour gradients \citep{La-Barber2012a, Kuntschner2010}, the radius to which stars are stripped within a galaxy determines the metallicity, and hence colour, of the stars that are liberated. 
Further into the cluster potential, tidal forces can reach deeper into a galaxy to strip redder, more metal-rich stars. 
This trend creates a radial colour gradient in the intracluster stellar population.

\item Major Mergers: A significant fraction of stars may be liberated from a galaxy merging with the BCG via violent relaxation \citep{Murante2007, Conroy2007, Lidman2013a}. 
These violent events serve to erode any existing stellar population gradient \citep{Kobayashi2004a, Di-Matteo2009a, Eigenthaler2013a}.
Thus, if central major mergers are a dominant channel for ICL formation then we should see relatively uniform ICL colour profiles.

\end{enumerate}

Fundamentally, the formation of the ICL depends on the types of galaxies and their interactions at the group or cluster centre.
Given the overabundance of early type galaxies in the centres of groups and clusters \citep{Dressler1980, Park2009a}, the dominant progenitor population and formation mechanism of the ICL is likely one that involves early type galaxies -- partial tidal stripping of massive galaxies and violent relaxation after central major mergers.
However, precisely which progenitor population is accountable for the majority of the ICL build-up remains uncertain. 
The observational results of \cite{Morishita2016} suggest that \logmstell$<$9.5 galaxies are the dominant contributor to the ICL.
Others recent observational studies favor galaxy-galaxy tidal interactions and tidal stripping via the cluster potential of \logmstell$<$10.5 galaxies for the origin of the ICL \citep{Annunziatella2016, Montes2014, Giallongo2014}. 

The complete disruption of lower mass satellites cannot be completely discounted as a means for ICL growth however.
Dwarf galaxies experiencing strong cluster tides or galaxy-galaxy interactions will invariably be disrupted and their stars will be added directly to the ICL. 
Recently, \cite{Annunziatella2016} have looked at the distribution of orbits of the dwarf galaxies in Abell 209.
They find a deficit of dwarfs with radial, plunging, orbits, which is consistent with the picture of either cluster tides or merging events with the central BCG as the dominant modes of ICL formation.
However, the fractional amount of the ICL contributed by these disrupted dwarfs is expected to be far less significant than that from tidal striping of moderate luminosity galaxies \citep{Contini2013a}.

The transfer of stars from galaxies to ICL will leave a mark on the luminosity function of galaxies in the cluster core. 
\cite{Giallongo2014} find that the luminosity function within 200 kpc of CL0024+17 exhibits a significantly shallower faint-end slope compared to  a composite luminosity function of galaxies out to the virial radius for clusters of similar redshift.
This difference in faint-end slope can be explained by a significant fraction of the stellar mass in intermediate and low-mass galaxies being removed via tidal interactions over time. 
Further, they compute the difference in emissivity between the inner and outer luminosity functions to be in the same range of their measured ICL luminosity fraction of \til23\%.  

Similarly, \cite{Annunziatella2016} look to differences in the stellar mass  function of the inner cluster compared to that of the entire cluster as a means to identify the ICL progenitor population.
In the case of Abell 209, they too find that the stellar mass function in the central region shows a deficit of galaxies at masses M$<$10$^{10.5}$\Msun, and that integrating over this `missing mass' adds up to the observed ICL mass.
They conclude that 90\%\ of the ICL in Abell 209 is consistent with originating in galaxies with 10$^{9-10}$ \Mstar.

In our own pilot study of the ICL in \cite{DeMaio2015} (hereafter Paper I), we find that the dominant formation mechanism of the ICL is likely tidal stripping of the outskirts of galaxies with L$>$0.2 \Lstar\ (\logmstell$>$10.4).
We disfavor central major mergers as a dominant formation mechanism of the ICL based on the observed blue-ward colour gradients, which cannot be produced via major mergers \citep{La-Barber2012a}.
Additionally, we find a total luminosity of ICL that is inconsistent with the expected frequency of violent major mergers in the formation history of the cluster since z$=$1 \citep{Lidman2013a}. 
Of the recent observational studies identifying the progenitor population of the ICL, ours (Paper I and this work) imposes the highest mass limit on the dominant contributors to the ICL build-up. We note that the models of \cite{Contini2013a} suggest that a significant fraction of the ICL originates in massive (\logmstell
$>$10.5) galaxies as well.

An unanswered question is if the dominant formation mechanism of the BCG+ICL changes for halos of different masses.
In lower-mass groups dynamical friction timescales are too long for mass segregation to occur \citep{Ziparo2013} and thus galaxy groups do not have the enhancement of massive galaxies at the group centre from which to build-up the ICL. 
However interaction times between galaxies are longer in galaxy groups, allowing for more efficient stripping via tidal processes. 
The number of high-to-intermediate-mass galaxies, those that likely contribute most to the ICL, are few, and thus the specific quantity of ICL in galaxy groups can vary widely, depending on the accretion history of the group. 
In particular, \cite{Contini2013a} model the ICL of low mass ($ 10^{13.4} < {\rm M}_{200}\ [{\rm M}_\odot] < 10^{13.6}$) haloes; they found larger ICL mass fractions for groups with relatively few, massive galaxies (${\rm M} >  10^{10} {\rm M}_\odot$) and smaller ICL mass fractions for groups with many more, less massive galaxies.

In this paper we expand on our results of Paper I by applying the same reduction and analysis techniques to produce radial surface brightness and colour profiles of the ICL to a maximum radius of  53-110 kpc for clusters from the \lclash\ (CLASH) survey \citep{Postman2012} with z$>$0.25 and 7 galaxy groups from \emph{Hubble Space Telescope (HST)} Program \#12575.
Together, these systems represent a sample of intermediate redshift clusters (0.29$\leq$z$\leq$0.89) with \mbox{\mfive\footnote{\mfive\ is the mass of a cluster within a radius where the cluster over-density is equal to 500 times the critical density of the Universe at the cluster redshift} from 3\tenthirteen\ to 9\tenfourteen\ \Msun.}
This study allows us to look to how a halo's mass affects the observed characteristics of its ICL.
In Section \ref{sec:reduction} we describe the reduction process, similar to that of Paper I with additional improvements to the flat-fielding and PSF subtraction. 
We present surface brightness profiles in Section \ref{sec:sb}.
colour profiles and how we derive ICL colour gradients are presented in Section \ref{sec:colorgrads}. 
In \S\ref{sec:lum_colordist} we discuss how the ICL luminosity and colour gradients behave as a function of halo mass as well as compare observed ICL colours to equivalent red sequence galaxy colours.
Throughout we use WMAP9 cosmology \citep{WMAP9}.

\label{sec:Intro}

\section{Sample}
Our sample consists of a combination of CLASH clusters \citep{Postman2012a} and galaxy groups from \hst\ Program \#12575.
The CLASH survey is a multi-cycle Treasury Program in which 25 clusters at 0.19$\leq$z$\geq$0.89 were imaged in 16 filters with Wide Field Camera 3 (WFC3)/ultraviolet and visible light (UVIS), WFC3/infrared (WFC3/IR) and Advanced Camera for Surveys (ACS)/Wide Field Camera (WFC).
CLASH reached 100\% completion in Cycle 20 after 524 orbits.
The CLASH sample consists of massive clusters ranging 
in mass from \mfive = 2.4\tenfourteen\ $-$ 9\tenfourteen\ \Msun. 
For our science goals we focus on the near-infrared \blue, \green, and \red\ filters (corresponding to broad bandpasses centred on 1.055\AA,1.153\AA, and 1.536\AA, respectively) from WFC3/IR for the subsample of 20 CLASH clusters with z$>$0.25.
The lower redshift limit is driven by the field of view of WFC3/IR.
At z$<$0.25, a 200 kpc distance is $>$50\arcsec, which leaves insufficient off-source area for sky determination. 

At the lower-mass end of our sample, we have seven groups with \mfive$<$1\tenfourteen\ \Msun.
Four of the galaxy groups are part of the supercluster SG1120 \citep{Gonzalez2005b}. 
Of the remaining groups, two are from the XMM Cluster Survey (XMM-XCS) survey \citep{Mehrtens2012} and two are from the ROSAT Deep Cluster Survey \citep{Mulchaey2006a}.
All of the galaxy groups have only a single orbit of imaging in \blue\ and \red, for a total of 8 orbits for the entire program. 
These groups were chosen because of their X-ray coverage as well as their intermediate redshift, which is well matched to the median redshift of $<$z$>$=0.4 of the CLASH sample.
Their redshift range also insures that there is sufficient area to use for a robust background subtraction and that the field of view (FOV) of WFC3 covers an appreciable fraction of \rfive\ (60-90\%).

\begin{figure}
\centering
\includegraphics[width=0.5\textwidth]{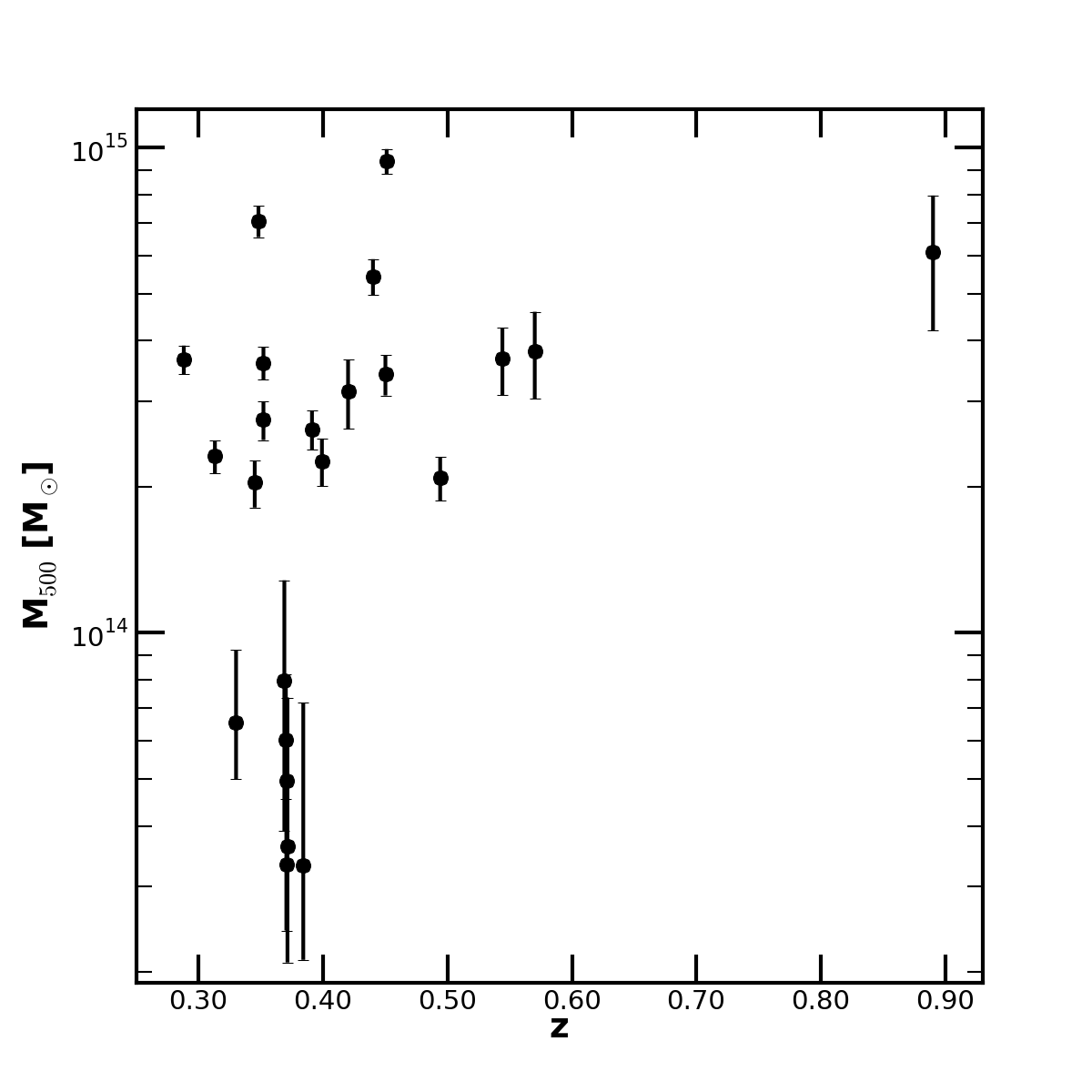}
\caption{\mfive\ vs. redshift, as derived by converting from X-ray temperatures using the \cite{Vikhlinin2009} prescription for the entire sample. See Table \ref{table:sample} for a breakdown of imaging and X-ray temperature references. This sample of 23 groups and clusters spans \mfive = 3\tenthirteen\ to 9\tenfourteen\ \Msun, making it the largest \hst\ sample of intermediate redshift galaxy groups and clusters used to study the ICL. }
\label{fig:sample}
\end{figure}

X-ray temperatures for all systems are sourced from the literature, which we use to determine \mfive\ masses using the \cite{Vikhlinin2009} prescription. 
For the CLASH clusters we use X-Ray temperature values from \chandra\ observations published in \cite{Postman2012a}.
For XMM022045 and XMM011140 we adopt X-Ray temperatures from \cite{Mehrtens2012}, which are based on \XMM\ data.
The X-ray temperature of RXJ1334 is also from \XMM\, and originates in \cite{Jeltema2006}.
Finally, the four groups of the super group SG1120 have X-Ray temperatures from \chandra, as in \cite{Gonzalez2005b}.
We are aware that cluster masses derived from \chandra\ are generally larger by \til15\% than those from \XMM\ X-Ray temperatures \citep{Mahdavi2013}.
However clusters with k$_B<$5 keV generally do not suffer this systematic difference \citep{Mahdavi2013} and thus we do not apply any corrective factor to the \mfive\ values derived for our group sample. 
The X-ray temperatures used to find \mfive\ for the CLASH clusters all originate from \chandra\ data, and thus do not need any corrections. 
Table \ref{table:sample} provides details for our sample, including their redshift, \mfive, and X-ray temperatures.
We show the distribution in mass and redshift of our sample in Figure \ref{fig:sample}.

\begin{table*}
\caption{Cluster Sample and Sources}
\centering
\begin{threeparttable}[b]
\tabcolsep=0.11cm
\small
\begin{tabular}{l l l l l l l}
Fullname & Cluster & z & kT & M$_{500}$ & r$_{500}$ & \hst, X-ray \\
 &  &  & [keV] & [10$^{14}$ M$_\odot$] & [kpc] & Source \\
\hline \hline
Abell611 & A611 & 0.288 & 7.9$\pm$0.35 & 3.66$\pm$0.25 & 996$_{-23}^{+22}$ & CLASH, a \\
MS2137-2353 & MS2137 & 0.313 & 5.9$\pm$0.3 & 2.31$\pm$0.18 & 847$_{-23}^{+21}$ & CLASH, a \\
XMMXCS J022045.1-032555.0 & XMM022045 & 0.330 & 2.6$_{-0.4}^{+0.7}$ & 0.65$_{-0.15}^{+0.27}$ & 552$_{-47}^{+67}$ & HST\#12575, d \\
RX J1532+3021 & RXJ1532 & 0.345 & 5.5$\pm$0.4 & 2.04$\pm$0.23 & 803$_{-31}^{+29}$ & CLASH, a \\
RX J2248-4431 & RXJ2248 & 0.348 & 12.4$\pm$0.6 & 7.06$\pm$0.52 & 1213$_{-31}^{+29}$ & CLASH, a \\
MACS1931-2635 & MACS1931 & 0.352 & 6.7$\pm$0.4 & 2.75$\pm$0.25 & 885$_{-28}^{+26}$ & CLASH, a \\
MACS1115+0129 & MACS1115 & 0.352 & 8.0$\pm$0.4 & 3.60$\pm$0.28 & 968$_{-25}^{+24}$ & CLASH, a \\
SG 1120-1202-4 & SG1120-4 & 0.369 & 3.0$_{-1.0}^{+1.2}$ & 0.8$_{-0.41}^{+0.49}$ & 582$_{-123}^{+100}$ & HST\#12575, b \\
XMMXCS J011140.3-453908.0 & XMM011140 & 0.370 & 2.5$_{-0.4}^{+0.6}$ & 0.6$_{-0.15}^{+0.22}$ & 530$_{-47}^{+58}$ & HST\#12575, d \\
SG 1120-1202-2 & SG1120-2 & 0.370 & 1.7$_{-0.3}^{+0.5}$ & 0.33$_{-0.09}^{+0.15}$ & 435$_{-43}^{+57}$ & HST\#12575, b \\
SG 1120-1202-1 & SG1120-1 & 0.371 & 2.2$_{-0.4}^{+0.7}$ & 0.49$_{-0.14}^{+0.24}$ & 496$_{-51}^{+70}$ & HST\#12575, b \\
SG 1120-1202-3 & SG1120-3 & 0.371 & 1.8$_{-0.5}^{+1.2}$ & 0.36$_{-0.15}^{+0.37}$ & 448$_{-75}^{+118}$ & HST\#12575, b \\
RX J1334.0+3750 & RXJ1334 & 0.384 & 1.7$_{-0.4}^{+1.3}$ & 0.33$_{-0.12}^{+0.39}$ & 432$_{-60}^{+127}$ & HST\#12575, c \\
MACS1720+3536 & MACS1720 & 0.391 & 6.6$\pm$0.4 & 2.63$\pm$0.24 & 859$_{-27}^{+26}$ & CLASH, a \\
MACS0429-0253 & MACS0429 & 0.399 & 6.0$\pm$0.44 & 2.26$\pm$0.25 & 815$_{-32}^{+29}$ & CLASH, a \\
MACS0416-2403 & MACS0416 & 0.420 & 7.5$\pm$0.8 & 3.14$\pm$0.51 & 902$_{-52}^{+47}$ & CLASH, a \\
MACS1206-0848 & MACS1206 & 0.440 & 10.8$\pm$0.6 & 5.43$\pm$0.46 & 1074$_{-31}^{+30}$ & CLASH, a \\
MACS0329-0211 & MACS0329 & 0.450 & 8.0$\pm$0.5 & 3.41$\pm$0.33 & 916$_{-30}^{+28}$ & CLASH, a \\
RX J1347-1145 & RXJ1347 & 0.451 & 15.5$\pm$0.6 & 9.38$\pm$0.56 & 1283$_{-26}^{+25}$ & CLASH, a \\
MACS1311-0310 & MACS1311 & 0.494 & 5.9$\pm$0.4 & 2.09$\pm$0.22 & 765$_{-27}^{+26}$ & CLASH, a \\
MACS1149+2223 & MACS1149 & 0.544 & 8.7$\pm$0.9 & 3.67$\pm$0.58 & 906$_{-51}^{+45}$ & CLASH, a \\
MACS2129-0741 & MACS2129 & 0.570 & 9.0$\pm$1.2 & 3.81$\pm$0.78 & 908$_{-67}^{+58}$ & CLASH, a \\
CL J1226+3332 & CL1226 & 0.890 & 13.8$\pm$2.8 & 6.08$\pm$1.89 & 937$_{-109}^{+88}$ & CLASH, a \\
\hline

\end{tabular}
\begin{tablenotes}[b]
    \item a: \cite{Postman2012a}, b: \cite{Gonzalez2005b}, c: \cite{Jeltema2006}, d: \cite{Mehrtens2012}
\end{tablenotes}
\end{threeparttable}
\label{table:sample}
\end{table*}

\section{Reduction}
\label{sec:reduction}

We follow the reduction methodology described in detail in Paper I with a few exceptions. In this section we summarise our reduction steps, going into greater detail for processes that differ from our methods in Paper I.
Since Paper I, we have refined the flat-fielding procedure and standardised the methodology for PSF subtraction for those fields with bright foreground stars. 

\subsection{Delta Flats}
\label{sec:deltaflat}
Because we are analyzing the very faint, diffuse ICL, we must take care that there is no large-scale residual variation in the flatness calibration of the WFC3/IR detector.
To this end we have created ``delta" flats for each filter by stacking WFC3/IR observations of sparse fields with exposure times between 100-1600s (typically several hundred images) bracketing the observation dates of these clusters. 
We apply these $\delta$-flats to all science images in addition to the pipeline flat-fielding with the flats of \cite{Pirzkal2011}.
After applying a 23x23 median smoothing kernel and a 5$\sigma$ iterative clipping of the large scale variations, we find the $\delta$-flats have rms differences from the Pirzkal flats of 0.8\%, 0.9\%, and 0.7\% in  \blue, \green, and \red, respectively.

We have made $\delta$-flats for each passband for a sequence of observation date ranges.
We multiply each calibrated individual exposure image (flt image) by its corresponding $\delta$-flat (See Table \ref{table:flat-epoch} for number of images input to each $\delta$-flat for each epoch.)
Flt images have been processed by \texttt{calwf3} (e.g. dark subtraction, bad pixel identification, flat-fielding, etc), but have not yet been drizzled into a final image (See WFC3 Handbook for more details). 
For \green, relatively few observations are available, and thus we have only created 2 $\delta$-flats.  
For the more commonly used \blue\ and \red\ we have created 3 $\delta$-flats.
Each flt image of our science images are matched to the $\delta$-flat with the  appropriate filter/epoch combination.
We show each epoch/filter $\delta$-flat in Figures \ref{fig:f105_delta-flat}- \ref{fig:f160_delta-flat}.

\subsection{\AD}
Since Paper I, \texttt{MultiDrizzle} has been replaced by \AD\ for the processing of \hst\ data from WFC3.
The most significant difference between our use of these two packages is that with \AD\ sky subtraction cannot be suppressed.
We allow \AD\ to run with sky subtraction so that it can properly identify cosmic rays, apply distortion corrections, and drizzle the flts images into the final science images.
To avoid over-subtraction of the background during drizzling we then create drizzled `sky frames'.
These frames are the result of drizzling the measured sky values taken from each input flt image in a given science image into a single, constant value at the same output pixel scale as the science images. 
These sky frames are added back to the drizzled science images.
We measure and subtract the background via custom methods at a later step, as described below. 

\subsection{Making the PSF}
\label{sec:psf}
In our previous analysis, we did not subtract foreground stars for two reasons: 1) no stars with \red$<$17 mag fall within the area in which we measured the ICL surface brightness and colour (r$<$ 100 kpc) and 2) bright stars outside of this radius could be masked while retaining a sufficient number of background pixels for the calculation of the background level.
All bright stars were masked to large radii to ensure that the light from the extended wings of the point spread function (PSF) was below our level of uncertainty in the measured background level.
With other CLASH and group fields we do not always have this luxury.
Thus, we must carefully account for the light in the extended halo of the PSF of bright stars because it may bias not only our background determination, but also contribute to artificially high ICL measurements. 

The model PSFs of TinyTim \citep{TinyTim} extend to a maximum of 15\arcsec, an insufficiently large radius for our purposes.
Further, only the inner 2\arcsec\ are recommended for use due to uncertainties in the models.
The profiles of the brightest stars in our images (\red\til14.5 mag) do not reach the uncertainty in the background until \til25".
Thus, we create a master PSF for each filter by identifying and stacking isolated, bright stars in several alternate fields (see Table \ref{table:psf_fields}) to increase the signal in the extended wings of the PSF.

To create the stacked, composite PSF for each filter, we first astrodrizzle each field in a manner identical to how we drizzle the science images.
Each image is masked of all objects but the star of interest and then normalised by the median flux in an annulus from 2.9-3.1\arcsec\ from the star centre. 
All stars from different fields for a given filter are stacked (excluding diffraction spikes).
We then construct a radial stellar profile composite.
By stacking bright stars we are able to derive a radial PSF profile out to \til28\arcsec, which ensures that we are completely subtracting any contribution of light in the extended wings from our ICL or background measurements for even the brightest stars.
We use this radial profile to construct a radially symmetric 2D PSF that is then used for the PSF subtraction in the science images. 

In each science field we identify stars brighter than \red=17.5 mag and then scale the composite 2D PSF to the observed star's brightness by performing a least square fit in which only the normalisation of the master PSF is variable. 
We then subtract the scaled 2D PSF from each bright star in the science image.
This PSF subtraction method ensures that the extended wings of foreground stars are not biasing our measurements and allow us to only mask the inner 5-8\arcsec\ of the brightest stars in the final science images, thus preserving as many pixels as possible for ICL and background measurement.

We also investigate the effect of the PSF on the convolution of the ICL profiles. 
If the extended wings of the PSF are different in each filter, it is possible that the convolution of the PSF with the BCG+ICL profile may artificially induce a colour gradient. 
To test the effect of the differential extended PSF wings, we first fit the surface brightness distribution of the BCG+ICL in \red\ of MACS1149.
Using a single component sersic profile fit centred on the BCG we find best-fit values of $\mu_e=$20.6, r$_e=$116 kpc, n$=$6.3.
We then generate an image with these parameters that has the same size and pixel scale as the original field and produce a radially averaged surface brightness profile from this image.
This model is then convolved with the 1D PSF profiles out to 26\arcsec\ in \blue, \green, and \red. 
If convolution with the PSF has no differential effect on the derived surface brightness profiles in these filters then we should recover a constant difference when we look at the colour of the convovled profiles.
Indeed, we find that the convolved colour profiles show a maximal difference of 0.01 \sbu\ from the expected constant colour outside of 0.5\arcsec.
We conclude that the convolution of the PSFs for the WFC3/IR filters used in this paper with the BCG+ICL surface brightness profiles does not impact our measured colour profiles.

\subsection{Masking}
\label{sec:masking}
\begin{figure*}
    \centering
    \includegraphics[width=\textwidth, trim={0, 3cm, 0, 3cm}, clip] {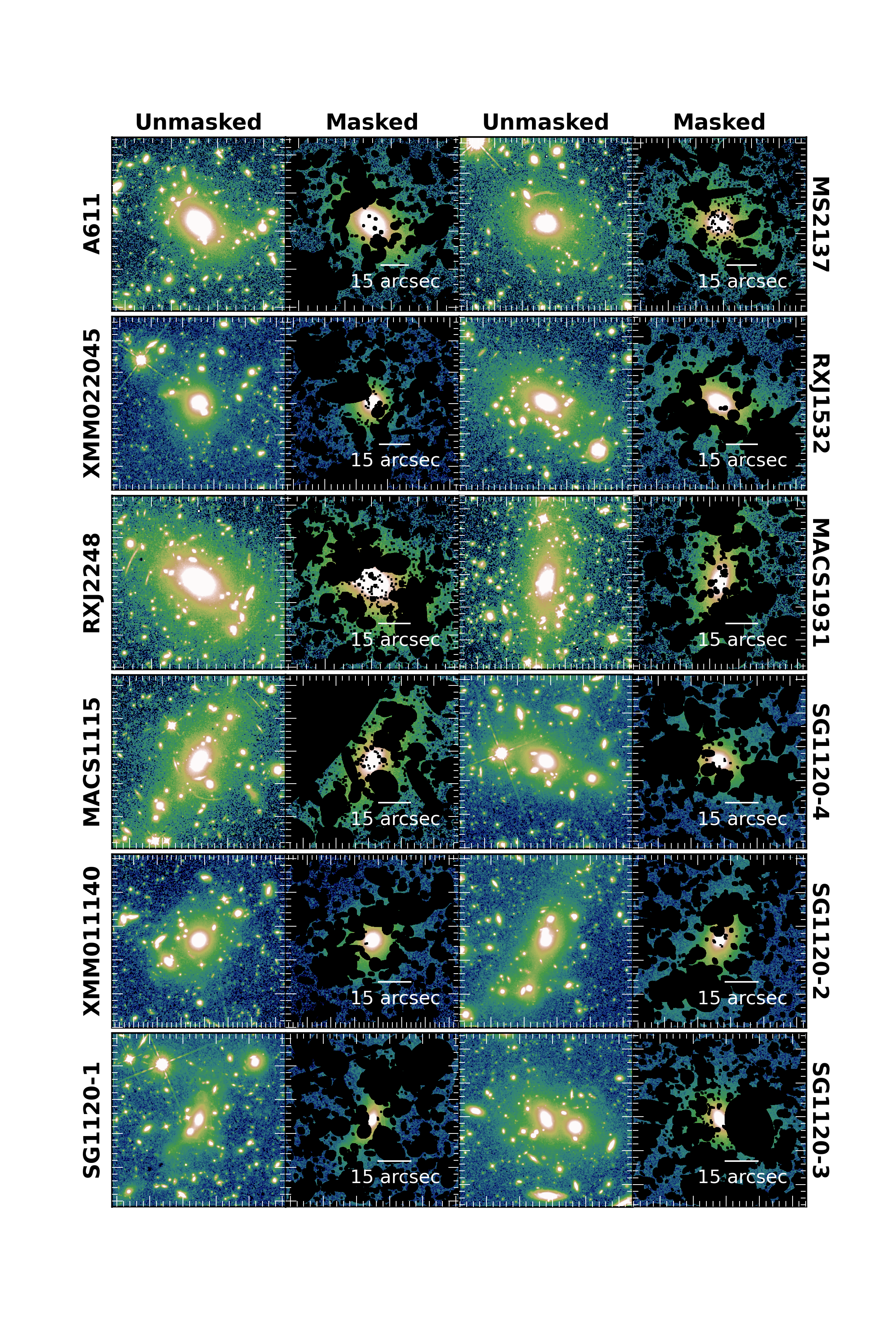}
    \caption{The inner 200 kpc of the masked and un-masked \red\ images for the twelve lowest redshift groups and clusters. In the unmasked images pixels brighter than 22 \sbu\ are in white and black regions in the masked images are masked sources. A 15\arcsec\ scale bar is marked on the masked cutouts. North is up, East left. } 
    \label{fig:postage_1of8}
\end{figure*}

\begin{figure*}
    \centering
    \includegraphics[width=\textwidth, trim={0, 3cm, 0, 3cm}, clip] {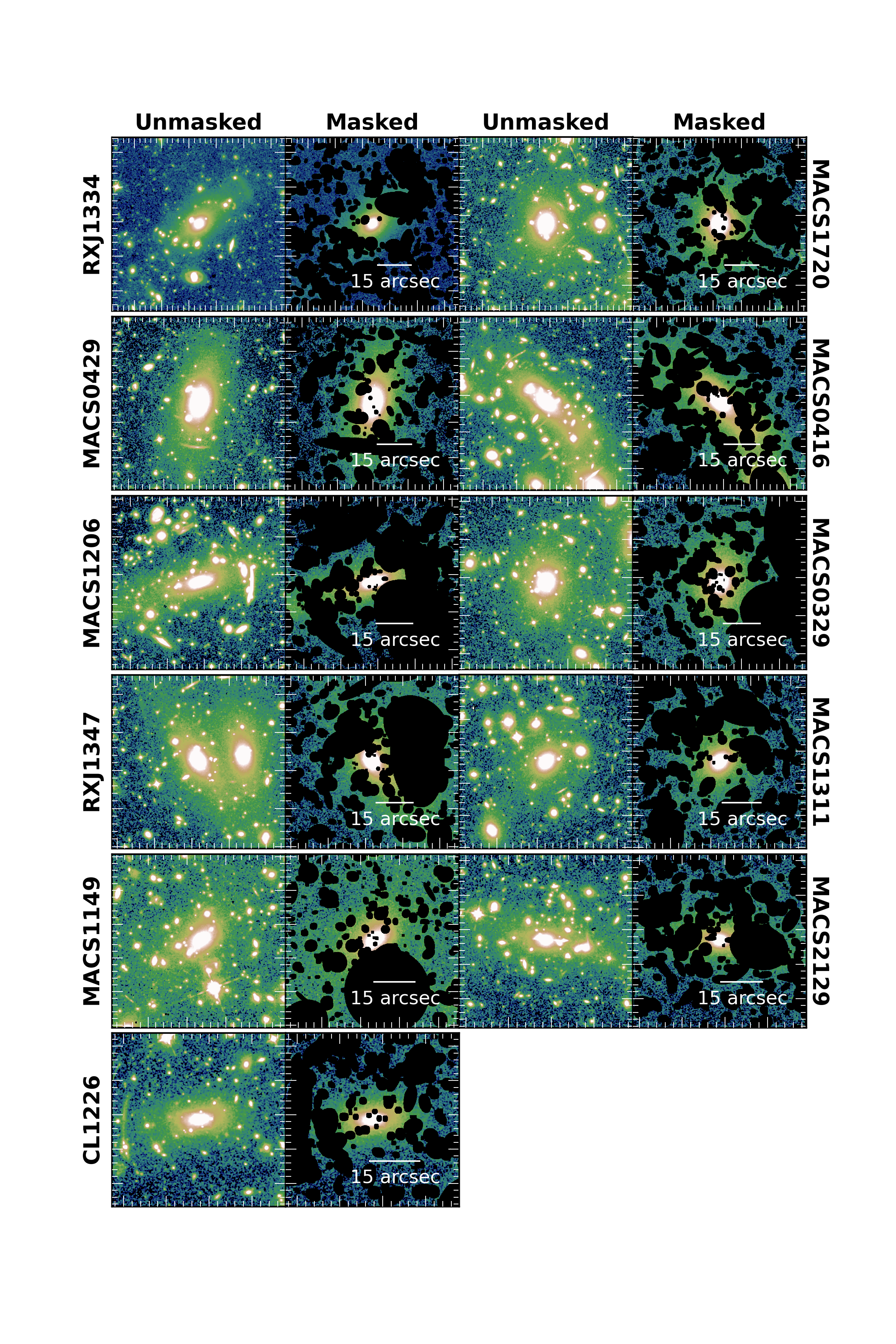}
    \caption{Same as in Figure \ref{fig:postage_1of8} but for the 11 highest redshift clusters. }
    \label{fig:postage_8of8}
\end{figure*}

After applying a $\delta$-flat to all input images and drizzling, we run Source Extractor \citep{SEx} for each group and cluster.
We use the F160W filter as the detection image to identify all sources.
After identification, we mask all sources more than 10\arcsec\ from the BCG to 3 times the semi-major and semi-minor axis output of Source Extractor, which is 2.5 times the Kron radius.
Within 10\arcsec\ of the BCG centre we manually mask sources by  extending the mask radius to eliminate galaxy contamination from the final ICL data.
Each epoch of data is masked individually and masked images are used to find the sky level.
The final mask for each filter is the combination of all the masks from all epochs of that filter. 
Finally, for a given colour (\blue$-$\red\ or \green$-$\red), we produce a final mask that combines the masks of both filters to insure that there is absolute symmetry in the masking and ICL extraction. 
See Figures \ref{fig:postage_1of8}-\ref{fig:postage_8of8} for pre- and post-masking examples of the inner 200 kpc of each cluster.

As discussed in Paper I, our results are robust to the masking method employed. 
What is important is that our masking is sufficiently extensive that the results are convergent rather than sensitive to the extended halos of individual galaxies. 
The analyses here use a fixed expansion factor of 3 times the output semi-major and semi-minor axis from Source Extractor to define mask sizes.
This methodology is similar to that employed by \cite{Jee2010} and \cite{KrickI}.
We also perform a test in which we vary the expansion factor between two and four in our current analysis. 
We find that both the surface brightness and colour profiles remain consistent within 1 sigma in all cases.
	
In addition to masking individual sources in each field, we also visually inspect images for any large-scale background structure changes between epochs and mask any significant, large-scale features.
While we account for residual large-scale features in the flatness of each image with our application of $\delta$-flats, similar features can be introduced into the observations from time-dependent sources such as scattered light.
We identify these phenomenon by taking difference images between epochs in a given filter. 
If we find large-scale features in these difference regions, we mask them out.
Masked pixels are not replaced or used in any measurement of ICL characteristics.
An instance of such masking can be seen for MACS1115 in Figure \ref{fig:postage_1of8}. 

\subsection{Background Subtraction}
\label{sec:skysub}
Because the parallel field observations associated with the CLASH clusters were not taken simultaneously with the science images, we cannot use the parallel fields to determine the background level of each cluster (See discussion in Paper I).
Thus, the background level must be determined from the science images themselves.
Our dominant source of systematic uncertainty is our background level measurement, which is impacted by the number of available sky pixels in each field that we can use to determine the background. 

To find the sky level in each epoch of data, we first excise a 300 (250) kpc radius circle centred on the BCG of each cluster with z$>$0.35 (z$<$0.35).
To have enough background pixels for a robust background determination, we only excise the inner 250kpc for clusters with z$<$0.35, as at that redshift 250 kpc corresponds to $>$50\arcsec, or nearly the full radius of the WFC3/IR detector.
We then fit and subtract a plane to all unmasked pixels beyond 300 (250) kpc. 

To test whether the ICL at 250 kpc and beyond appreciably elevates the observed background level, we take the surface brightness profiles in \red\ and fit them with a simple Sersic model extending to large radii ($>$300 kpc). 
The integrated ICL light from 300-450 kpc (roughly the largest radius inside of which we determine the sky) corresponds to a flux that is always below the 1$\sigma$ uncertainty level of the background, often by as much as  by 1-2 \sbu.
Thus even for ICL colours at 200 kpc, a radius to which we generally are not able to measure the colour due to the background uncertainty at such surface brightness levels (\til27 \sbu\ in \red), we expect minimal systematic bias from using pixels at 250 kpc and beyond in our background measurement. 

After excising the inner 250-300 kpc, we then divide the remaining unmasked pixels into twenty-four, fifteen degree wedges centred on the BCG.
For each wedge we perform a 3$\sigma$ iterative clip. 
The final sky value used for that epoch is then the mean of all wedge values, taking the standard error in the mean as the background level uncertainty. 
See Table \ref{table:sky} for the background value and error for each epoch of data for all clusters.

\section{Surface Brightness Profiles}
\label{sec:sb}
After masking, background gradient subtraction, PSF subtraction (if necessary), and sky subtraction, we measure the median radial surface brightness profile of the ICL in each filter.
Masked pixels in each annuli are ignored and not replaced.

We use decadic logarithmic bin widths to maintain roughly equal signal-to-noise in each radial bin.
We measure the ICL surface brightness in dlog(r[kpc])=0.05 and dlog(r[kpc])=0.15 bins, taking the median radius value in each bin as the bin radius.
The \dlogr=0.05 bins are narrow, and thus the derived surface brightness profiles show considerable bin-to-bin scatter.
However, because the bins are so narrow, there is less ambiguity in the bin radius.
We use these profiles in subsequent radial trend fits. 
The \dlogr=0.15 bins produce smoothed profiles with less scatter, which we use for clarity of visual representation in this paper's figures.

Because of the multi-epoch imaging available with CLASH we can constrain the systematic variation in our surface brightness measurements by comparing the measured surface brightness of a given cluster on different observation dates. 
We take this systematic uncertainty, evaluated as the scatter in surface brightness measurements for a given cluster and filter over the available observation dates, as the error in our surface brightness measurements because the statistical errors are sub-dominant to the systematic uncertainty.
However, because we have only a single epoch of \blue\ and \red\ data for the galaxy groups, we cannot constrain the groups' surface brightness measurements in the same way. 
Instead, we make a composite error on the surface brightness to surface brightness ($\sigma_\mu-\mu$) relation of all CLASH clusters for \blue\ and \red\ and fit each relation with an exponential function. 
To find the uncertainty on the groups' surface brightness measurements in a given filter we then use the best-fit exponential and take the corresponding uncertainty given the observed surface brightness. 
This proceedure assumes that the variation in surface brightness measurements from the multi-epoch imaging of the CLASH survey also represents the level of systematic error measuring the surface brightness of the single-epoch galaxy group images. 
The data for the groups was acquired in the same time frame as the CLASH sample. 

\begin{figure*}
    \centering
    \includegraphics[width=\linewidth]{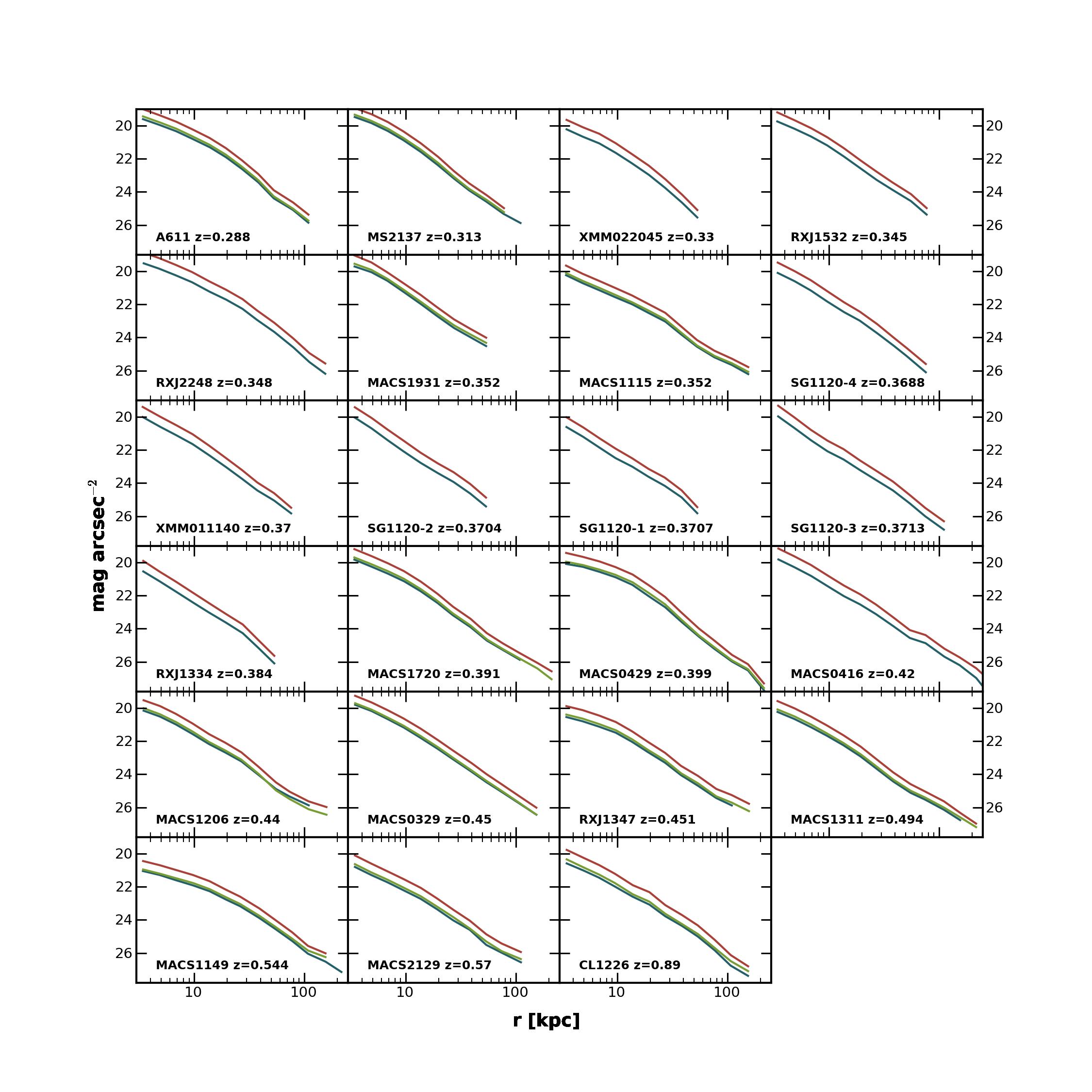}
    \caption{Surface brightness profiles for \blue, \green, \& \red\ corresponding to the blue, green, and red lines, respectively. Clusters and groups are ordered by increasing redshift. Profiles are terminated when 3 consecutive bins in the \dlogr=0.05 bins have $>$0.2 \sbu\ uncertainty in the measured surface brightness. Depending on filter and cluster, this criteria is reach at \til26-27 \sbu\ for the CLASH clusters and \til25-26 \sbu\ for the groups. Error bars are omitted as they are generally too small to be seen and are always less than 0.2 \sbu.}
    \label{fig:all_sbprofs}
\end{figure*}

In Figure \ref{fig:all_sbprofs} we show the observed (no passband, evolution, or cosmological dimming corrections) \blue, \green, \& \red\ surface brightness profiles (blue, green, and red lines, respectively) for all clusters and groups ordered by redshift, left to right.
Groups and three of the CLASH clusters are limited to \red\ and \blue\ profiles (as described in \S\ref{sec:reject} below).
All systems' surface brightness profiles are truncated when 3 consecutive bins in the \dlogr=0.05 profiles have an uncertainty greater than 0.2 \sbu\ (\til20\% relative uncertainty). 
The radius at which this criteria is reached varies cluster to cluster, but generally lies in the range of 26-27 \sbu\ for the CLASH clusters.
Because the uncertainty in \blue\ is higher and we have only a single orbit in both \blue\ and \red\ for the galaxy groups, their profiles reach this criteria at smaller radii, which corresponds to a brighter surface brightness limit. 

\section{Radial colour Profiles}
\label{sec:colorgrads}
We produce ICL radial colour profiles by subtracting the \red\ surface brightness profiles from either the \blue\ or \green\ surface brightness bin by bin.
In the very core region, the low numbers of pixels in each bin could result in colour profiles that differ from those derived if we instead first produced a difference image and then radially binned. 
However, at r$>$10 kpc low numbers of pixels in each bin is not a concern and we do not use the colour profiles within 10 kpc in any quantitative way throughout the extent of this paper. 
The same masks are used for both filters, ensuring that any observed features in the ICL are physical and not artifacts of different masking.

Our dominant uncertainty is in the measurement of the background, which varies by as much as 75\%, 59\%, and 45\% between epochs in \blue, \green, and \red, respectively.  
To constrain the systematic errors between different filters and epochs we analyze each epoch of data separately.
This means that for the CLASH clusters we have an ensemble of 8 colour profiles (all CLASH clusters have 2 \green\ epochs and 4 \red\ epochs - for a total of 8 possible combinations.)
We use the spread between all the individual epochs of colour profiles to constrain the systematic errors between different epochs of data. 
These systematic uncertainties are considerably larger than any intrinsic uncertainties for a given bin.
We therefore take the error in the mean of all individual colour profiles per bin as the error in the final colour of the ICL in that bin. 

As a measure of the robustness of our ICL colour profiles we compare the average colour profile produced with each of the bluer filter epochs (e.g. the average of all four \red\ image subtracted from a single \green\ epoch image).
\begin{equation}
\textrm{F110W}_i - <\textrm{F160W}> = \frac{\Sigma^{N_{\textrm{F160W}}}_{j=1}( \textrm{F110W}_i - \textrm{F160W}_j)}{N_{F160W} }
\end{equation}
A CLASH cluster is only included in our final sample if these two profiles of a given colour (either \blue$-$\red\ or \green-\red) are consistent within 2$\sigma$ at all radii (see \S\ref{sec:reject}).

We find that differences between the two epochs of the bluer filter (either \blue\ or \green) are the largest source of systematic uncertainty in the colour profiles.
This is because the bluer filters are more affected by scattered light and HeI emission from the upper atmosphere \citep{HSTHeI}.
Further, the background noise in \green\ is less affected by the diffuse emission of HeI in the upper atmosphere because of its wide wavelength coverage as compared to \blue's smaller bandwidth. 
For most clusters we are able to minimise any large-scale structure difference in the two epochs of the bluer filter observations by looking at the difference image of the two epochs and masking out holes, peaks, or other large-scale structures in the background (as described in \S \ref{sec:masking}).

For the groups we employ a similar technique to assess the \blue$-$\red\ profiles produced. 
The \blue\ images of the groups consist of 6-8 calibrated individual exposure images (flt image) which are drizzled into the final science images.
To assess the spread in measured colour we split the single orbit of \blue\ data into 2 sets of 3-4 flt images and drizzle them as if they were two different epochs of data. 
As with the CLASH clusters, the spread in the `multi-epoch' group colour profiles is used to constrain the robustness of our measured ICL colours. 
Half of the flts from SG1120-3 and SG1120-4 are contaminated with HeI emission, and were removed.
Thus, for these two groups we are unable to split the \blue\ images into artificial epochs.
The final colour profiles of both SG1120-3 and SG1120-4 are the single epoch colours with uncertainties reflecting only the systematic surface brightness measurements uncertainties from the composite CLASH relationship, as described in \S\ref{sec:sb}. 

\subsection{Rejected Groups and Clusters}
\label{sec:reject}

In general, for both groups and CLASH clusters, the \blue$-$\red\ colours show larger uncertainty due to \blue\ systematically suffering from higher background uncertainty.
Of the 20 CLASH clusters with z$>$0.25, 13 yield robust \green$-$\red\ profiles.
Three of the clusters without solid \green$-$\red\ profiles do, however, have robust \blue$-$\red\ profiles (RXJ1532, MACS0416, and RXJ2248). 
To bring the groups and the three CLASH clusters with \blue$-$\red\ profiles to a common colour, we apply a colour correction. 
See Appendix \S\ref{sec:color-corr} for the details of this transformation.

\begin{figure*}
\centering
\includegraphics[width=\textwidth]{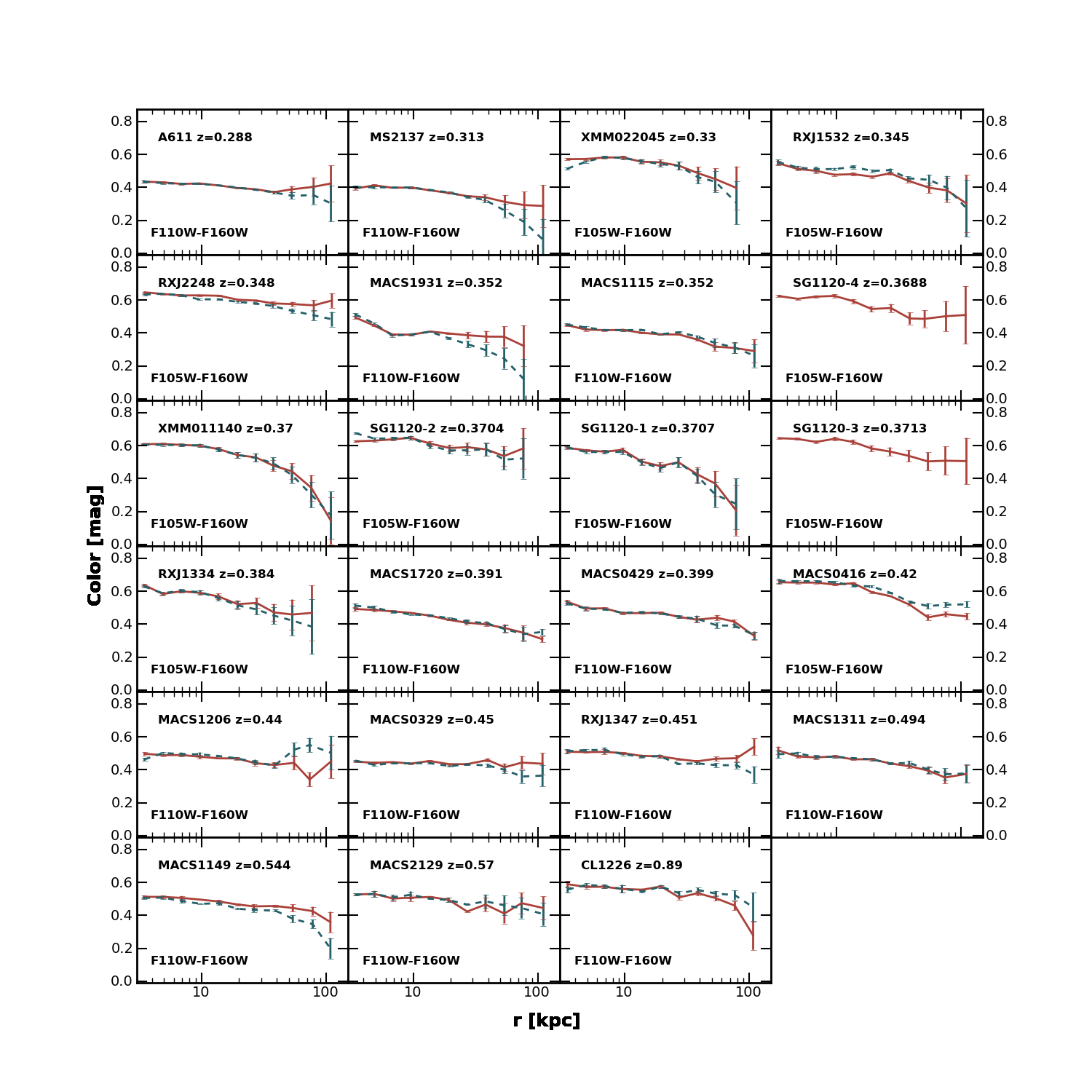}
\caption{ICL colour profiles off all systems, ordered by redshift left to right. \green$-$\red\ is shown for all CLASH clusters except for MACS0416, RXJ2249, RXJ1532 and the groups. For these systems we use the \blue$-$\red\ colour profiles of the ICL and convert to \green$-$\red\ using a colour correction derived from a BC03 solar metallicity, SSP model generated with \ezgal\ (See \S\ref{sec:color-corr}). For systems with multiple epochs of the bluer filter, two colour profiles are shown, one for each profile produce by averaging all \red\ images with a single bluer filter image. Profiles are terminated at the point where the uncertainty in the colour becomes larger than 0.2 \sbu\ in 3 consistent bins in the \dlogr=0.05 radial profiles. }
\label{fig:color_panel}
\end{figure*}

Figure \ref{fig:color_panel} illustrates the raw colour profiles of each system in the final sample that were used to assess whether to include a cluster in the final sample.
Of the 20 CLASH clusters available, we exclude MACS1423+2404, MACS0647.8+7015, and MACS0744.9+3927 because the average colour profiles produced with each epoch of their bluer filter (as in Figure \ref{fig:color_panel} for the final sample) are of poor data quality over the majority of the radial range probed in both \blue$-$\red\ and \green$-$\red.
Similarly, we exclude RXJ0329 from the final group sample based on inconsistent \blue$-$\red\ profiles beyond \til20 kpc. 
Finally, we exclude MACS0717 from the final sample because it is a very dynamic systems of 4 merging clusters with no clear central BCG to which to anchor the radial profiles \citep{Limousin2012}. 

\subsection{Measuring colour Gradients}
\label{sec:radprofs}

\begin{figure*}
    \centering
    \begin{subfigure}{0.48\textwidth}
    \centering
        \includegraphics[width=\textwidth]{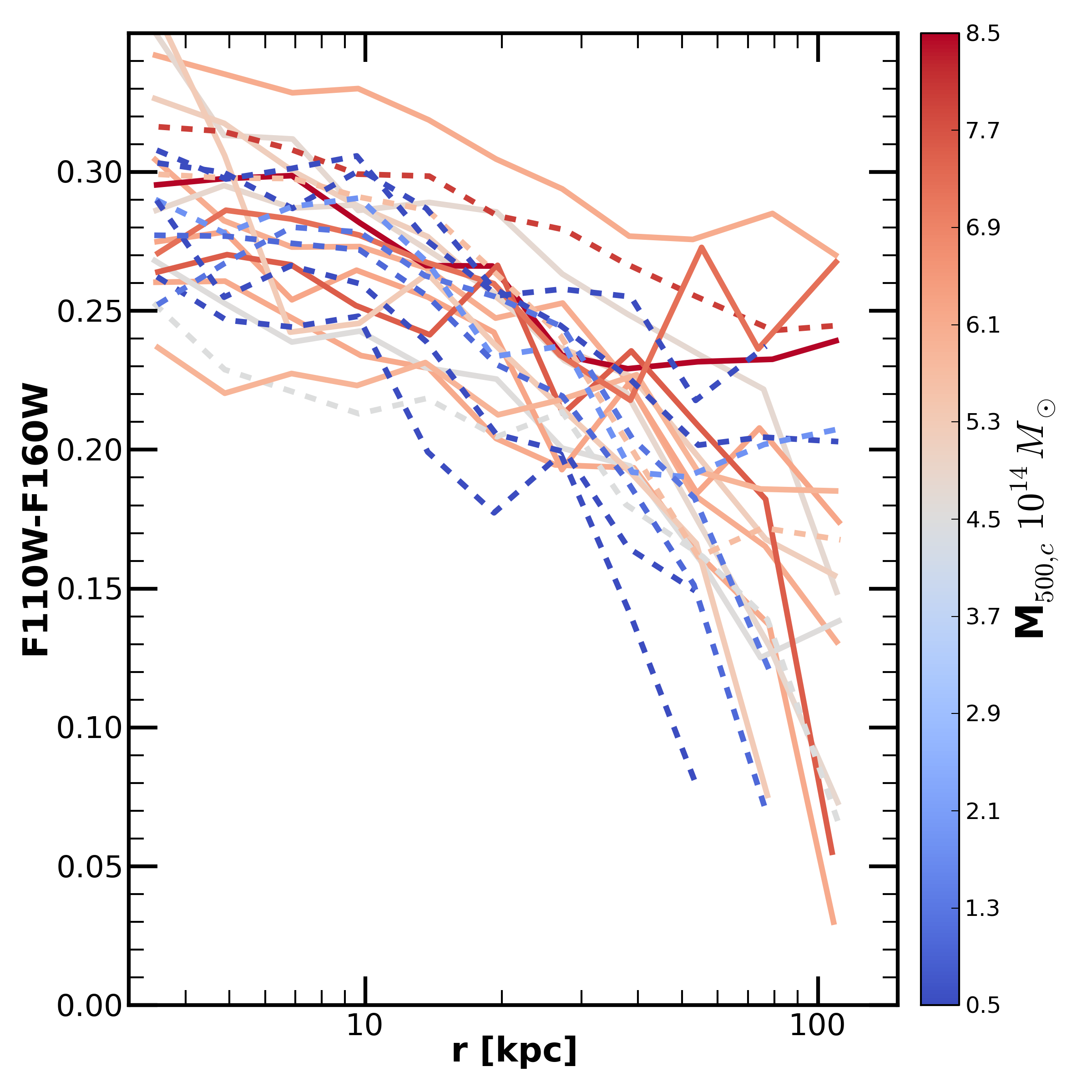}
        \caption{ }
        \label{fig:stacked}
    \end{subfigure}
    \begin{subfigure}{0.48\textwidth}
    \centering
        \includegraphics[width=\textwidth]{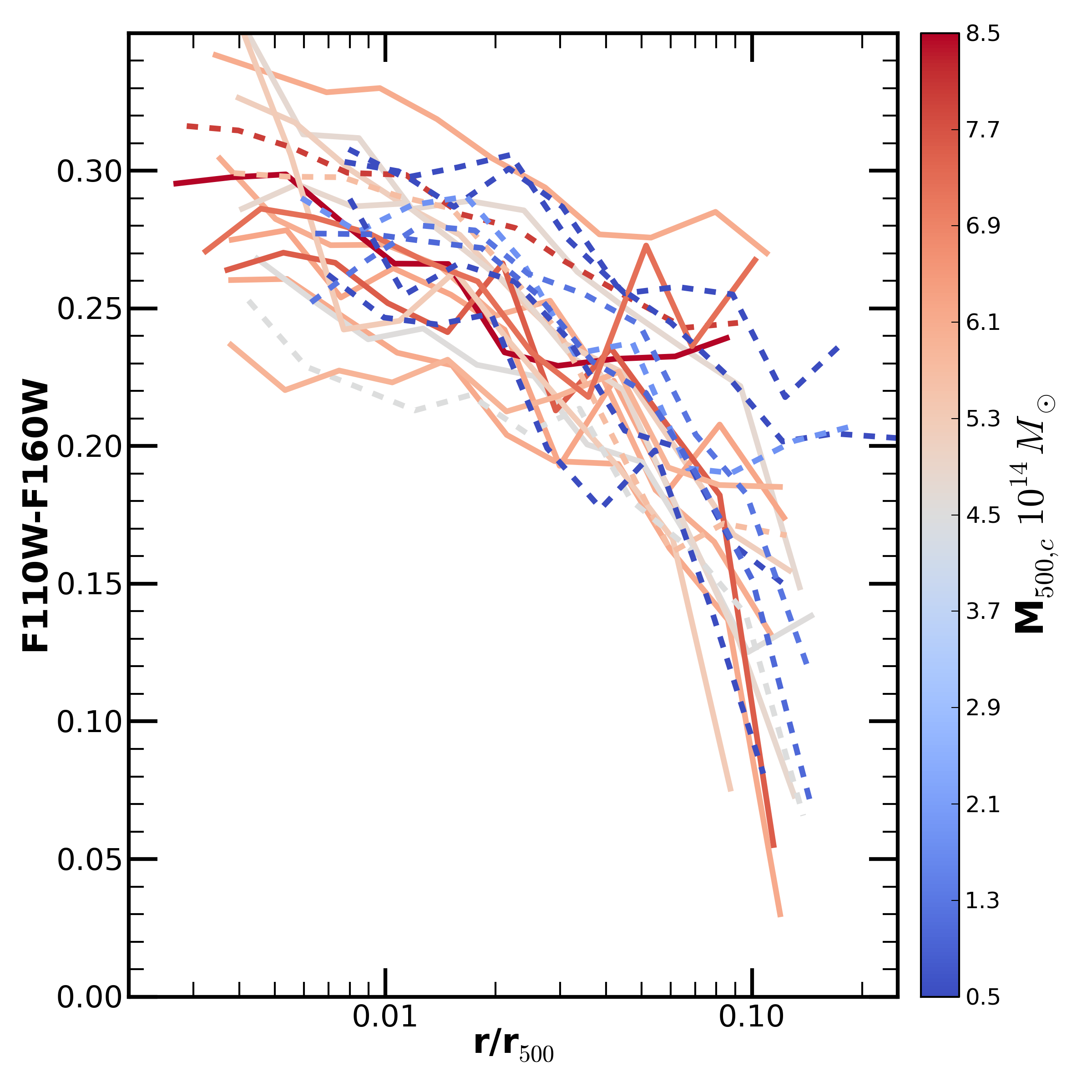}
        \caption{ }
        \label{fig:stacked_r500}
    \end{subfigure}
    \caption{\green$-$\red\ colour profiles of all systems. Systems observed in \blue$-$\red\ (dashed lines) have been corrected to \green$-$\red\ colours as described in \S\ref{sec:color-corr}. \emph{Left: } All profiles are e+k corrected to z$=$0. \emph{Right: } \green$-$\red\ colour profiles scaled by each system's \rfive. All systems show a similar shape in their colour profile. They are generally flat inside of 10 kpc, where the BCG dominates, and gradually become bluer with increasing radius. The negative colour gradients indicate that tidal stripping and dwarf disruption are the likely dominant formation mechanisms of the ICL. 
    }
\end{figure*}

To measure the colour gradient of the ICL, we combine all colour epochs into a single average profile for bin sizes of both dlog(r[kpc]) = 0.05 and 0.15, taking the median radius in each bin as the bin location.
For the dlog(r[kpc])=0.05 binned profiles we do not terminate the profiles until there are 3 consecutive bins that have a colour uncertainty of greater than 0.2 \sbu. 
Clusters reach this point at various radii -- the groups extend to 53-120 kpc and the CLASH clusters reach between 75-250 kpc. 
In Figure \ref{fig:stacked} we show \green$-$\red\ colour profiles e+k corrected to z$=$0 for our entire sample of systems. 
We use \ezgal\ \citep{ezgal} to estimate the e+k corrections for a \cite{BC03} (hereafter BC03) simple stellar population model with solar metallicity, formation redshift of \zform$=$3, and Chabrier initial mass function (IMF) \citep{Chabrier2003}.
We have omitted error bars for clarity, representative uncertainties are shown in Figure \ref{fig:color_panel}.
Figure \ref{fig:stacked_r500} shows the same profiles as in Figure \ref{fig:stacked} but each system is scaled to its \rfive.
A tabular version of the observed (not e+k corrected) \green$-$\red\ profiles in \dlogr=0.15 bins, along with 1$\sigma$ errors, are available in the appendix (See Table \ref{table:tabular_color}).

We measure the ICL colour gradients using the \dlogr=0.05 radial bins, and define the colour gradient, \colorgrad, as $\nabla_{F110W-F160W}$ henceforth.
The radial extent of the colour profiles varies significantly between groups and clusters.
If we fit all systems to a constant outer radius, then this restricts us to a maximum radius of 53 kpc (RXJ1334). 
Conversely, requiring data extending to 110 kpc would eliminate 6 of the 7 group systems.
To understand the effect of maximum radius on the measured colour gradient, we first fit all systems to 53 kpc. 
For those systems whose colour profiles reach 110 kpc we then refit the colour gradient.
We find that no systems are significantly affected by changing the outer radius of the fit from 53 to 110 kpc, though fits using the larger outer radius have lower uncertainties.

We next bin our systems into clusters (\mfive$>$1\tenfourteen\ \Msun) and groups (\mfive$<$1\tenfourteen\ \Msun) perform a simultaneous fit to all profiles in each bin to compute a best fit ensemble gradient. 
For the groups we find a colour gradient of \nab=$-$0.143$\pm$0.025. 
For the CLASH clusters we find \nab=$-$0.093$\pm$0.011.
At face value, these fits argue that the group and cluster gradients differ at the 2$\sigma$ level.

To test the robustness of this result, we next assess the impact of the choice of inner fitting radius.  
In the discussion thus far we have been using a nominal inner radius of 10 kpc. 
The choice of inner radius is potentially important because we must avoid the central region where the BCG dominates the observed luminosity.
In this regime, the BCG colour gradients are typically much flatter than for the outer ICL profiles. 
Inclusion of radii for which the BCG dominates the luminosity will thus bias the observed measurements towards shallower ICL colour gradients. 
We repeat the joint fit using an inner radius of 15 kpc. 
With this revised inner radius we find \nab=$-$0.114$\pm$0.040 for groups and \nab=$-$0.104$\pm$0.015 for the clusters--values that are consistent to within the observational uncertainties. 
We thus see no statistically significant evidence for a mass dependence of the observed colour gradients.

We also recompute the individual profiles using 15 kpc for the inner radius. 
The colour gradients of the ICL remains in the range of -0.25$\leq$\nab$<$0 for all systems whether they are measured with an inner fitting radius of 10 or 15 kpc. 
For individual systems, particularly the groups, values can change by more than the statistical uncertainties. 
These changes reflect the fact that the highest signal-to-noise comes from the smaller radii, so any colour variations due to structure on these scales can impact the overall fits. 
Despite the ambiguity in colour gradient, which depends on the choice of inner fitting radius, we find that all but one cluster show a negative colour gradient at the $\ge$3$\sigma$ level.
The outlier, MACS0329, was one of the four systems studied in Paper I, where we first noted its flat colour gradient.
As a check, we compare the best-fit colour gradients of the four clusters from Paper I to those derived with the re-processing described in this Paper. 
For these four clusters (MACS1206, MACS0329, MACS1149, \& MACS2129) we find best-fit slopes that are consistent with the published values of Paper I within 1.5$\sigma$.

We also perform a simultaneous fit to all colour profiles to find the characteristic colour gradient for the entire sample of groups and clusters.
Using a $\chi^2$ minimization, we fit for the colour gradient that best represents the entire sample by simultaneously fitting all colour profiles from 10$<$r$<$110 kpc, letting the normalisation of the fit vary for each cluster. 
We find an ensemble best-fit slope of \nab=$-$0.097$\pm$0.012.
Repeating this evaluation for all colour profiles with an inner radius of 15 kpc we find \nab=$-$0.105$\pm$0.018, which is consistent within 1$\sigma$ of the best-fit using a 10 kpc inner cutoff.

We express the colour gradients in terms of physical units (kpc), but the logarithmic definition of the gradient (dlog(flux)/dlog(radius)) means that a simple scaling of radii by \rfive\ will not affect the quantitative measure of the gradient.
The potential effect of considering objects that span a range of sizes arises only from scale-dependent deviations from a power-law gradient between groups and clusters, such as the role of the BCG that we explored with the selection of the inner radius. 
Ultimately, we selected to express the gradients in physical units because that is the simpler, most robust expression that can be compared to simulations.

\section{ICL Luminosity and colour Gradient Distribution}
\label{sec:lum_colordist}
In Paper I we had a much smaller sample of clusters -- only 4  CLASH clusters.
Still, based on the observed negative colour gradients combined with the high ICL luminosities of those first four clusters we concluded that tidal stripping of L$>$0.2\Lstar\ galaxies is the dominant means by which the ICL builds up.
With this expanded sample consisting of 16 massive CLASH clusters and 7 less-massive galaxy groups, we have a greatly expanded sample with which to test if these conclusions hold, particularly as we can investigate the effect of total cluster mass on ICL characteristics. 

Of the 23 groups and clusters in our sample, all but one show negative colour gradients at the 3$\sigma$ level or higher (best-fit gradients are listed in Table \ref{table:best_fit}).
From our simultaneous fits to the binned group and cluster colour profiles we find that the groups have a characteristic colour gradient ranging between \nab=$-$0.143$\pm$0.025 to \nab=$-$0.114$\pm$0.040, for colour profiles measured with an inner radius of 10 and 15 kpc, respectively. 
For the more massive clusters the best-fit colour gradient ranges from \nab=$-$0.093$\pm$0.011 to $-$0.104$\pm$0.015.

Such negative colour gradients can be produced by either dwarf disruption or tidal stripping, but not via violent relaxation after major mergers with the BCG \citep[See \S \ref{sec:Intro}]{Eigenthaler2013a, La-Barber2012a, Melnick2012}. 
We conclude that either tidal stripping or dwarf disruption are the dominant channels of ICL growth over a wide range of cluster masses for systems at z$<$0.9. 
However, we cannot discriminate between the two mechanisms using the observed negative colour gradients alone. 

\begin{sidewaystable*}
\caption{colour Gradients, Luminosity, and Stellar Mass of BCG+ICL}
\begin{threeparttable}
\begin{tabular}{ccccccccccc}
\hline
Cluster & z & $\nabla_{F110W-F160W}$ & Best-fit radius & L (r$\le$10 kpc) & L (r$\le$50 kpc) & L (r$\le$100 kpc) & M$_\bigstar$ (r$\le$10 kpc) & M$_\bigstar$ (r$\le$50 kpc) & M$_\bigstar$ (r$\le$100 kpc) \\
 &  & [\sbu log(kpc)$^{-1}$] & [kpc] & [$10^{11} L_\odot$] & [$10^{11} L_\odot$] & [$10^{11} L_\odot$] & [$10^{11} M_\odot$] & [$10^{11} M_\odot$] & [$10^{11} M_\odot$] \\
 \hline
A611 & 0.288 & -0.075$\pm$0.004 & 110 & 6.59$\pm$0.02 & 20.21$\pm$0.17 & 24.69$\pm$0.77 & 3.42$\pm$0.01 & 10.49$\pm$0.09 & 12.81$\pm$0.40 \\
MS2137 & 0.313 & -0.129$^{0.013}_{0.012}$ & 106 & 6.96$\pm$0.02 & 15.93$\pm$0.22 & 17.60$\pm$0.45 & 3.62$\pm$0.01 & 8.28$\pm$0.11 & 9.14$\pm$0.23 \\
XMM022045 & 0.330 & -0.123$^{0.026}_{0.004}$ & 67 & 4.28$\pm$0.02 & 9.45$\pm$0.09 & 10.54$\pm$0.30 & 2.23$\pm$0.01 & 4.92$\pm$0.05 & 5.48$\pm$0.16 \\
RXJ1532 & 0.345 & -0.060$^{0.015}_{0.013}$ & 106 & 6.00$\pm$0.04 & 15.21$\pm$0.27 & 17.05$\pm$0.54 & 3.12$\pm$0.02 & 7.92$\pm$0.14 & 8.88$\pm$0.28 \\
RXJ2248 & 0.348 & -0.053$\pm$0.002 & 110 & 8.47$\pm$0.03 & 30.82$\pm$0.45 & 35.26$\pm$0.95 & 4.41$\pm$0.01 & 16.05$\pm$0.23 & 18.36$\pm$0.49 \\
MACS1931 & 0.352 & -0.096$\pm$0.002 & 75 & 6.19$\pm$0.03 & 15.62$\pm$0.46 & 20.23$\pm$1.91 & 3.23$\pm$0.01 & 8.14$\pm$0.24 & 10.54$\pm$1.00 \\
MACS1115 & 0.352 & -0.111$^{0.006}_{0.005}$ & 110 & 3.88$\pm$0.02 & 13.65$\pm$0.32 & 18.50$\pm$1.36 & 2.02$\pm$0.01 & 7.11$\pm$0.16 & 9.64$\pm$0.71 \\
SG1120-4 & 0.369 & -0.144$^{0.004}_{0.003}$ & 106 & 4.89$\pm$0.02 & 11.22$\pm$0.10 & 13.12$\pm$0.33 & 2.55$\pm$0.01 & 5.86$\pm$0.05 & 6.85$\pm$0.17 \\
XMM011140 & 0.370 & -0.173$^{0.074}_{0.055}$ & 84 & 4.91$\pm$0.02 & 11.05$\pm$0.10 & 13.04$\pm$0.37 & 2.56$\pm$0.01 & 5.77$\pm$0.05 & 6.81$\pm$0.19 \\
SG1120-2 & 0.370 & -0.120$^{0.006}_{0.010}$ & 67 & 4.63$\pm$0.02 & 9.70$\pm$0.13 & 11.15$\pm$0.47 & 2.42$\pm$0.01 & 5.07$\pm$0.07 & 5.82$\pm$0.25 \\
SG1120-1 & 0.371 & -0.222$\pm$0.011 & 67 & 2.50$\pm$0.02 & 6.22$\pm$0.10 & 7.17$\pm$0.36 & 1.31$\pm$0.01 & 3.25$\pm$0.05 & 3.75$\pm$0.19 \\
SG1120-3 & 0.371 & -0.119$^{0.016}_{0.013}$ & 110 & 5.34$\pm$0.02 & 11.19$\pm$0.11 & 12.35$\pm$0.30 & 2.79$\pm$0.01 & 5.84$\pm$0.06 & 6.45$\pm$0.16 \\
RXJ1334 & 0.384 & -0.154$^{0.060}_{0.005}$ & 53 & 2.94$\pm$0.02 & 6.28$\pm$0.10 & 6.73$\pm$0.29 & 1.54$\pm$0.01 & 3.28$\pm$0.05 & 3.52$\pm$0.15 \\
MACS1720 & 0.391 & -0.125$^{0.003}_{0.002}$ & 110 & 6.74$\pm$0.02 & 17.20$\pm$0.32 & 21.70$\pm$1.25 & 3.52$\pm$0.01 & 8.99$\pm$0.17 & 11.35$\pm$0.66 \\
MACS0429 & 0.399 & -0.096$^{0.011}_{0.009}$ & 110 & 6.74$\pm$0.02 & 23.65$\pm$0.26 & 28.84$\pm$0.91 & 3.53$\pm$0.01 & 12.38$\pm$0.13 & 15.09$\pm$0.48 \\
MACS0416 & 0.420 & -0.152$^{0.028}_{0.021}$ & 110 & 6.40$\pm$0.02 & 18.33$\pm$0.44 & 22.43$\pm$0.85 & 3.35$\pm$0.01 & 9.60$\pm$0.23 & 11.75$\pm$0.45 \\
MACS1206 & 0.440 & -0.094$^{0.013}_{0.010}$ & 110 & 5.32$\pm$0.02 & 15.58$\pm$0.46 & 20.64$\pm$1.81 & 2.79$\pm$0.01 & 8.17$\pm$0.24 & 10.83$\pm$0.95 \\
MACS0329 & 0.450 & -0.013$^{0.012}_{0.011}$ & 110 & 6.77$\pm$0.02 & 19.09$\pm$0.44 & 25.21$\pm$1.78 & 3.55$\pm$0.01 & 10.02$\pm$0.23 & 13.23$\pm$0.93 \\
RXJ1347 & 0.451 & -0.085$^{0.015}_{0.012}$ & 110 & 5.53$\pm$0.02 & 16.84$\pm$0.51 & 22.89$\pm$2.10 & 2.90$\pm$0.01 & 8.84$\pm$0.27 & 12.02$\pm$1.10 \\
MACS1311 & 0.494 & -0.116$^{0.014}_{0.013}$ & 110 & 5.79$\pm$0.01 & 14.26$\pm$0.21 & 16.51$\pm$0.37 & 3.05$\pm$0.01 & 7.50$\pm$0.11 & 8.69$\pm$0.20 \\
MACS1149 & 0.544 & -0.123$^{0.043}_{0.034}$ & 110 & 3.59$\pm$0.02 & 17.64$\pm$0.65 & 23.95$\pm$2.61 & 1.90$\pm$0.01 & 9.32$\pm$0.34 & 12.65$\pm$1.38 \\
MACS2129 & 0.570 & -0.103$^{0.022}_{0.015}$ & 110 & 3.99$\pm$0.03 & 11.71$\pm$0.37 & 13.56$\pm$0.71 & 2.11$\pm$0.02 & 6.19$\pm$0.20 & 7.17$\pm$0.37 \\
CL1226 & 0.890 & -0.071$^{0.017}_{0.015}$ & 110 & 9.01$\pm$0.04 & 28.41$\pm$0.24 & 36.03$\pm$0.92 & 4.87$\pm$0.02 & 15.36$\pm$0.13 & 19.48$\pm$0.50 \\

\hline
\end{tabular}
\begin{tablenotes}
    \item Quoted uncertainties on the stellar mass do not reflect systematic errors associated with the mass to light conversion, which is model and IMF dependent. Nor does it account for any possible gradient in the M/L ratio of the ICL.
\end{tablenotes}
\end{threeparttable}
\label{table:best_fit}
\end{sidewaystable*}

As in Paper I, we use the total ICL luminosity and stellar mass to break the degeneracy between tidal stripping and dwarf disruption.
We convert the radially averaged \red\ flux profile into equivalent \Lsun\ luminosities.  
Values for the BCG+ICL luminosity within 10, 50, and 100 kpc are listed in Table \ref{table:best_fit}. 

For r$<$100 kpc we find total BCG+ICL luminosities within 100 kpc of  L$_{BCG+ICL}>$1.4\tentwelve\ \Lsun\ (14 \Lstar) for the CLASH clusters ($<$L$_{CLASH,100}>$=2.3\tentwelve\ \Lsun (23 \Lstar)) and between 0.7-1.3\tentwelve\ \Lsun\ ($<$L$_{GROUP,100}>$=1.1\tentwelve\ \Lsun, $>$11 \Lstar) for the groups 
(using \Lstar=1\teneleven\ \Lsun, as estimated with a BC03 \citep{BC03} model with Coma normalisation, Chabrier IMF \citep{Chabrier2003}, solar metallicity, and formation redshift \zform=3 at z$=$0.5).
These luminosities are too great to be explained by dwarf disruption alone -- shredding the number of dwarfs it would take to equal this luminosity would leave an indelible mark on the faint end slope of the member galaxy luminosity function.

On average the groups have a BCG+ICL luminosity between 10$<$r$<$100 kpc of L$_{10-100}=$6.4\teneleven\ \Lsun (\til6 \Lstar).
To produce this level of luminosity, the faint end slope of the galaxy luminosity function (GLF) would need to evolve from at least as steep as $\alpha$=-1.85 to present day values of $\alpha$ \til-0.8 \citep{Lin2004a,Muzzin2007}.
Such an extreme change in the faint population of galaxies in clusters is inconsistent with the observed lack of evolution in $\alpha$ since at least z\til1.3, and potentially since z\til3.2 \citep{Mancone2012a, Strazzullo2010, Wylezalek2014}.
Dwarf galaxies below the completeness limits of these studies are not the answer either; producing the observed ICL
luminosities from such faint galaxies alone would require a near-divergent GLF faint end slope.
Thus, we infer that dwarf disruption cannot account for the majority of the ICL regardless of the total mass of the cluster in which the ICL is being formed.

Further, the average amount of BCG+ICL luminosity observed between 10$<$r$<$100 kpc in the group sample, which is considerably less than that for the more massive CLASH clusters, cannot be explained exclusively by violent relaxation after major mergers with the BCG. 
If each major merger event deposits 20-50\% (as in \cite{Murante2007} or \cite{Lidman2013a}) of the incoming \Lstar\ galaxy into the ICL this would require 12-30 merging events to account for the average 6 \Lstar of light in the ICL (10$<$r$<$100 kpc) of the galaxy groups. 
This number of mergers is a few to 10 times higher than the expected number of major mergers after z=1 \citep{Lidman2013a}, which is when the majority of the ICL is expected to build-up. 

\begin{figure}
\centering
    \includegraphics[width=\columnwidth]{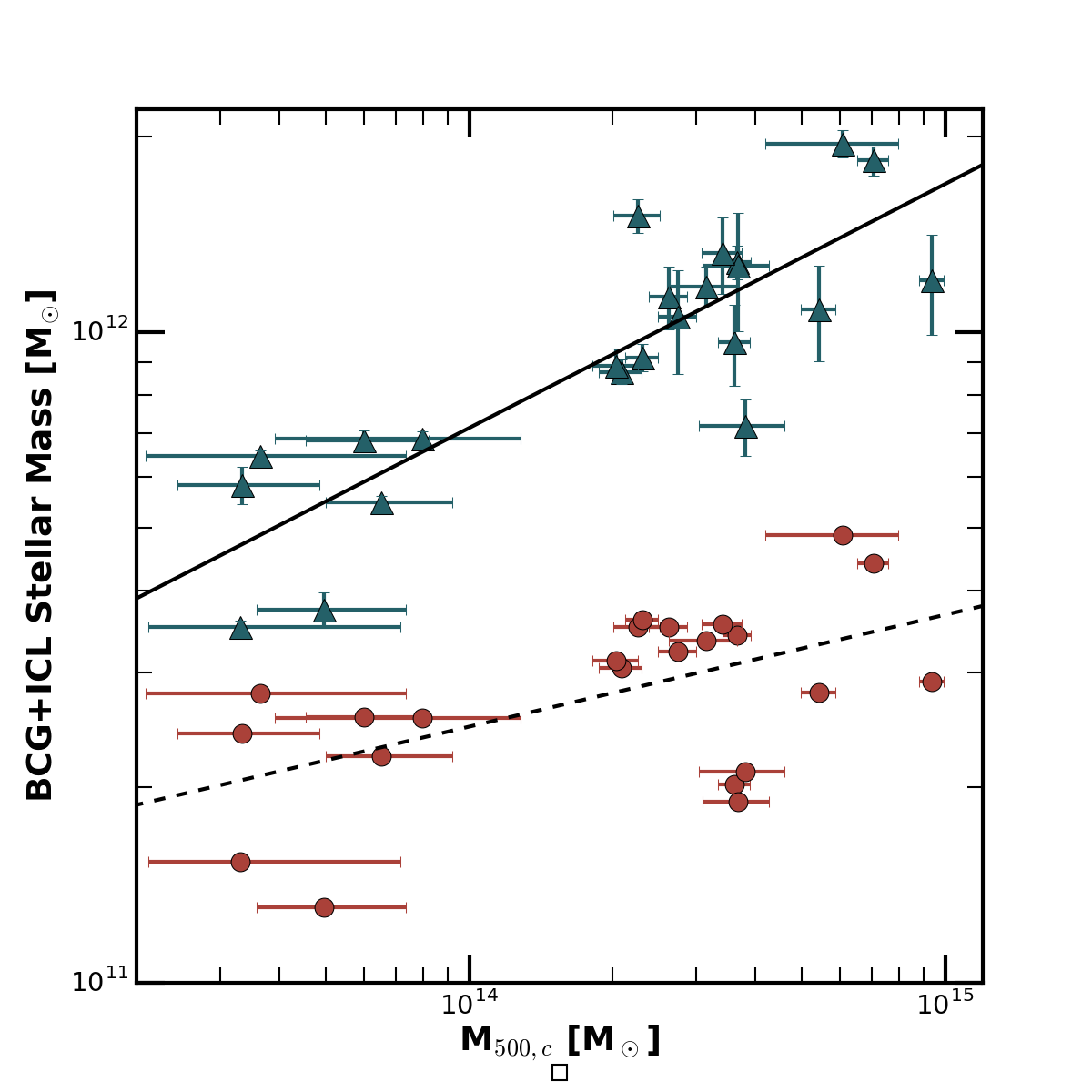}
\caption{BCG+ICL stellar mass as a function of \mfive. Circles show the total stellar mass for radii $<$ 10 kpc and triangles for r$<$100 kpc. The stellar mass in the inner r$<$10 kpc has a much shallower form as a function of \mfive\ (dashed line) with a fit log(\Mstell)$=$(0.17$\pm$0.06)\logmfive+(8.99$\pm$0.80), as compared to the total stellar mass within 100 kpc (solid line) with a fit log(\Mstell)$=$(0.37$\pm$0.05)\logmfive+(6.63$\pm$0.70). }
\label{fig:lums}
\end{figure}

In Figure \ref{fig:lums} we show the BCG+ICL stellar mass, \Mstell, as a function of \mfive\ in the inner 10 kpc (circles) and r$<$100 kpc (triangles) for each system.
We arrive at these stellar masses by taking our measured BCG+ICL luminosities and applying a solar mass to light ratio from a BC03 simple stellar population (SSP) model with Chabrier IMF, formation redshift \zform=3, and solar metallicity.
For the groups the stellar mass at r$<$10 kpc, which is dominated by the BCG, has a mean value of 2.2\teneleven\ \Msun.
For clusters with \mfive$>$1\tenfourteen\ \Msun\ the equivalent stellar mass is 3.2\teneleven\ \Msun.
At r$<$100 kpc, these values grow to 5.5\teneleven\ \Msun\ and 1.2\tentwelve\ \Msun\ for groups and clusters, respectively. 
The trend of increasing BCG+ICL stellar mass with total cluster mass is consistent with other recent baryon census studies and models of galaxy groups and clusters \citep{Gonzalez2013a, Contini2013a, Lagana2013, Zhang2007}.

We fit the \Mstell--\mfive\ relations with a power law for the total stellar mass within 10 and 100 kpc.
We find \Mstell$_{,10}\propto$\mfive$^{0.17\pm0.06}$ and \Mstell$_{,100}\propto$\mfive$^{0.37\pm0.05}$
The stellar mass within the inner 10 kpc has a shallower index (0.17$\pm$0.06) as compared to the r$<$100 kpc index (0.37$\pm$0.05).
This difference suggests that while BCG+ICL mass grows with cluster mass, there is a maximum stellar density threshold which prevents further central growth, even in the largest clusters. 
The total stellar mass instead must grow primarily from accretion at larger radii.

We note that these fits are to \mfive\ values that correspond to the mass of the clusters at their redshift (not evolved to z$=$0) and that our sample does not constitute an evolutionary sample.
Thus we cannot speak to how the BCG+ICL stellar mass distribution changes as a given halo evolves.
However, we can conclude that for a given cluster mass the ICL contains a larger fraction of the stellar content of the cluster core than the BCG (r$<$10 kpc) and that this inequality grows with increasing total mass. 
Additionally, we find that although the stellar content of the BCG+ICL goes up with total mass, it increases more slowly than the host cluster total mass.
Qualitatively this suggests that low-mass, group environments, are more efficient at producing ICL than clusters within a fixed physical radius.
These conclusions echo those found by \cite{Gonzalez2013a} for a sample of low-redshift clusters with 1\tenfourteen$<$\mfive$<$1\tenfifteen\ \Msun. 

\subsection{Equivalent Galaxy colours}
\label{sec:equivcolor}
Several recent observational studies use the ICL colour in comparison to the galaxy member population colours to constrain which galaxies contribute most to the build-up of the ICL \citep{KrickI, KrickII, Montes2014, Morishita2016}.
This can only be a rough comparison; only the lowest mass galaxies are completely disrupted and thus fully deposited into the ICL and therefore match the observed ICL colour one-to-one.

We use the observed colour of the ICL at each radius to constrain the galaxy progenitor population. 
In the simplest case, where a single type of galaxy dominates the contribution, then the colour of the ICL will match the colour of the progenitor population. 
In reality, we expect a variety of galaxies to contribute.
However, because the galaxy red sequence sets a limit on reddest galaxies possible and because that limit is a function of luminosity and therefore mass, one can constrain the progenitor population by requiring that the reddest galaxy contributing to the ICL not already be bluer than the ICL. 
For example, if the reddest low mass galaxy is bluer than the ICL, then low mass galaxies cannot be the dominant progenitor population of the ICL.
In this way we investigate how the dominant galaxy donor changes as a function of radius, and how more realistic physical scenarios, in which the donor red sequence galaxy is not completely disrupted, affect the range of galaxies that can significantly contribute to the ICL. 

\cite{Connor2017} have created a photometric catalog of all the CLASH fields, which uses multi-scale, mode-based background subtraction to detect both the full extent of large galaxies as well as small galaxies embedded in diffuse cluster emission across all 17 CLASH filters. 
colours are measured in fixed apertures between filters, at sizes comparable to the Kron radius, after subtraction of individual local backgrounds. 
They then calculate photometric redshift probabilities using the Baysian Photometric Redshift code presented by \citet{BPZ2000} \citep[BPZ\footnote{http://www.stsci.edu/ \til dcoe/BPZ/}][]{Benitez2004, Coe2006} and model stellar populations for each galaxy with \texttt{iSEDfit} \citep{Moustakas2013}. 
\cite{Connor2017} define a set of cluster members for each cluster, based on the following ordered priorities: spectroscopic redshift (where available), photometric redshift probability, SED goodness-of-fit, and a less stringent photometric redshift probability; a more detailed description of this selection is provided in that paper.

After we apply the CLASH member identification to the catalogs, we e+k correct the galaxy magnitudes to z=0.
For each cluster we then subtract m$_{*,F160W}$ from the measured \red\ galaxy magnitudes to normalise all clusters' members magnitudes relative to m$_*$.
We determine m$_{*,F160W}$ for each cluster using a Coma normalisation with a BC03 SSP with \zform=3.0, solar metallicity, and Chabrier IMF. 
We then create a composite colour-magnitude diagram in \green$-$\red\ and perform a weighted linear fit to determine the red sequence. 
We initially fit to all galaxies, and then clip galaxies beyond 0.15 mag from the best fit line and fit again. 
This relation allows us to translate the observed ICL colours to the magnitude of a red sequence galaxy with equivalent colour.
We find that the bluest ICL colours in the range of 10$<$r$<$100 kpc (0.008-0.13 \rfive) are consistent with cluster galaxy colours at \Mstar+2.5, equivalent to L$>$0.1 \Lstar, as seen in Figure \ref{fig:mstarx}.

We note that \Mstar+2.5 is only the \emph{equivalent} magnitude galaxy that matches the observed ICL colour at 100 kpc; this assumes red sequence galaxies, regardless of mass, are completely disrupted to build the ICL.
However, galaxies are not necessarily fully disrupted and the extent to which stars are removed depends on both the mass and orbit of the galaxy.
Because of the internal colour gradients in early-type galaxies, there is actually some range in colour of stars that can be stripped into the ICL.
For more massive galaxies in many cases only the outskirts are stripped during tidal interactions and therefore only stars bluer than the total integrated colour of the galaxy are added to the ICL.
Thus, our calculation of the mass of galaxies contributing to the ICL in the complete disruption scenario is a conservative lower limit on the progenitor population of the ICL as a function of radius.

\begin{figure}
\centering
    \includegraphics[width=\columnwidth]{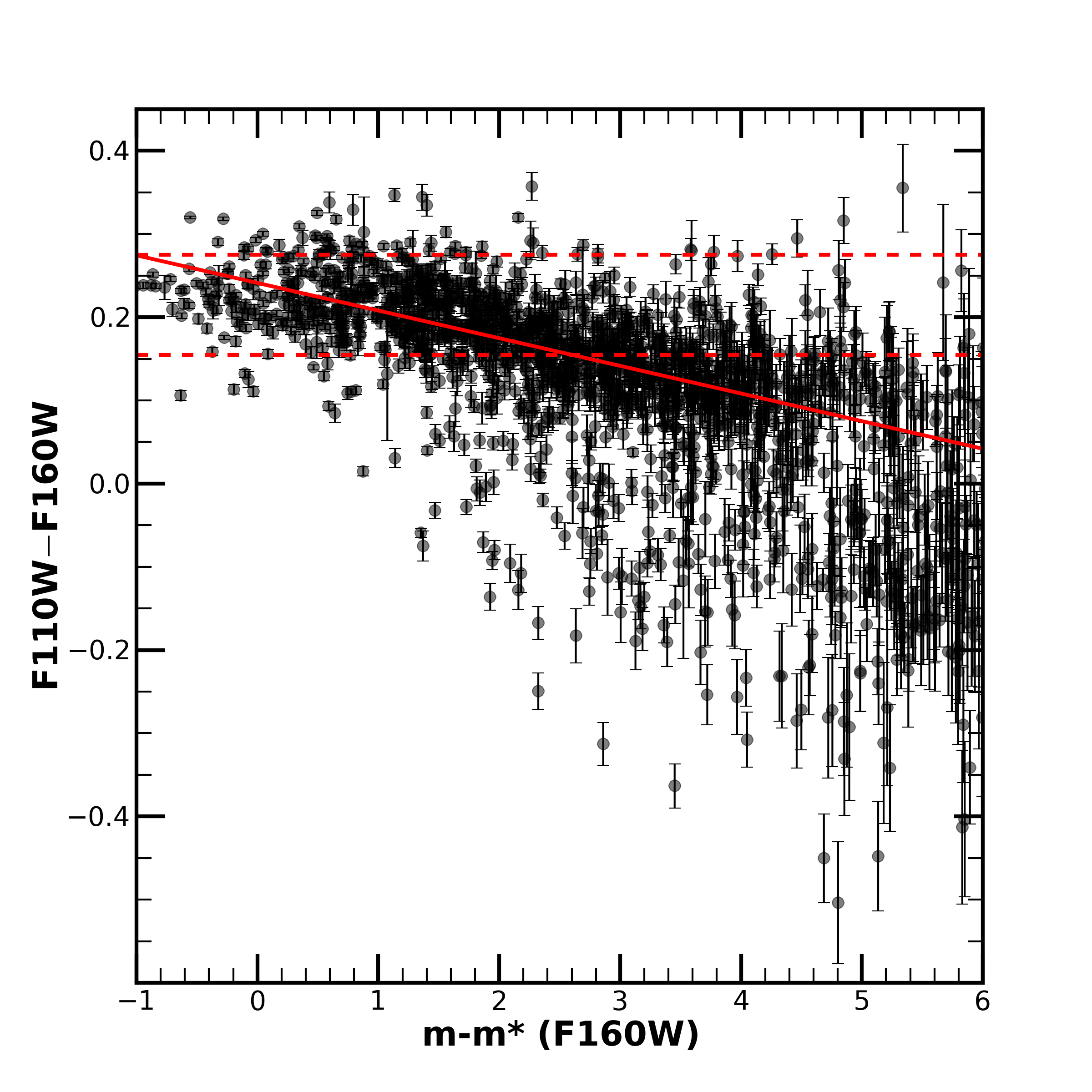}
    \caption{Composite CMD for all CLASH clusters, using cluster membership catalogs from \cite{Connor2017} (as described in \S\ref{sec:equivcolor}). Each cluster is e+k corrected to z=0 with \ezgal\ using a BC03 model with SSP, \zform=3, Chabrier IMF, and solar metallicity. We find m$_*$ for each cluster with the same SPS model with a Coma normalisation. We then normalise all \red\ magnitudes to m$_{*,\textrm{F160W}}$ by taking  m$_{F160W}$ - m$_{*,\textrm{F160W}}$ for all clusters. Our red sequence fit is illustrated with the solid red line. Horizontal dashed lines represent the median ICL CLASH colour at r=10 kpc (upper) and r=100 kpc (lower).}
    \label{fig:cmd}
\end{figure}

\begin{figure}
\centering
    \includegraphics[width=\columnwidth]{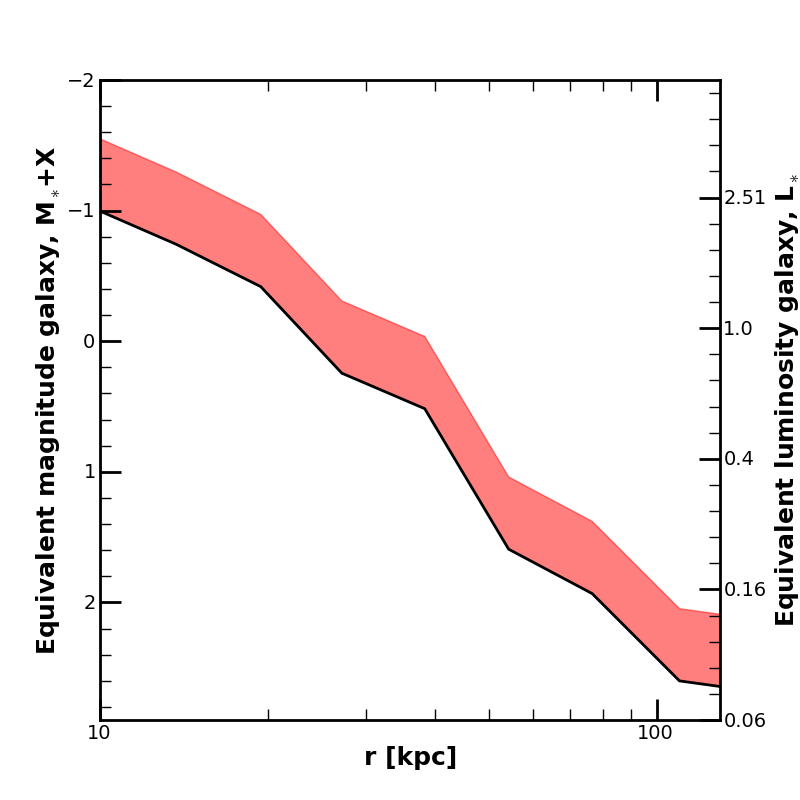}
    \caption{Using our best-fit red sequence as in Figure \ref{fig:cmd}, we translate the observed colour of the median CLASH ICL colour profile at each radius to a red sequence galaxy with an equivalent colour. In this way we estimate the progenitor population of the ICL \emph{if all galaxies that contribute to the ICL are completely disrupted}. The red shaded region represents the possible range when instead of complete disruption, galaxies are only stripped from 1-2 times the radius the drives the observed luminosity-weighted colour. The total disruption (black) line represents a hard lower-limit on the magnitude of galaxy that can significantly contribute to the ICL at that radius. A galaxy with the same equivalent colour as the ICL at 100 kpc (\til0.1\rfive) corresponds to a \Mstar+2.5 galaxy. }
    \label{fig:mstarx}
\end{figure}

From \cite{La-Barber2010a}, the internal colour gradient of massive early-type galaxies in the near-infrared is $\nabla_{Y-H}$=-0.061 mag/log(r).
Assuming stars are only stripped from 1-2 times the radius that drives the observed luminosity-weighted colour, this corresponds to stars that are 0.018 mag bluer than the integrated colour of the galaxy.
In order to match the observed ICL colour to these bluer stars we must, in effect, move the red-sequence relation (as in Fig \ref{fig:cmd}) bluer by 0.018 mag. 
We then repeat the conversion from observed ICL colour to equivalent galaxy magnitude with this shifted red-sequence.
The ambiguity in ICL progenitor source (total disruption vs. tidal stripping) is represented in Figure \ref{fig:mstarx} as the red shaded region, which encompasses galaxies \til0.5 mag brighter than the total disruption (solid line) scenario.

This procedure, as described in the preceding text, is only a first order estimation to the colour range of stars liberated from a galaxy's outskirts.
The actual tidal radius to which a galaxy is stripped depends not only on the galaxy mass, but also on the location within the potential at which it is being stripped.
However, using this metric, we can place constraints on the dominant ICL progenitor galaxy population as a function of cluster radius -- at a given radius there is typically a range of \til0.5 mag in the magnitude of galaxies contributing stars to the ICL assuming a stripping depth of 1-2 times the radius which drives the observed luminosity-weighted colour. 
Further, we can use the derived equivalent magnitude galaxy versus radius (Figure \ref{fig:mstarx}) in conjunction with the median colour profile versus radius of the CLASH clusters to identify the fraction of the total BCG+ICL luminosity within r$<$100 kpc as a function of cluster member galaxy magnitude.
In this way, we infer that \til75\% of the total BCG+ICL luminosity within r$<$100 kpc is consistent with red sequence galaxy magnitudes of \Mstar+1.6 and brighter (log(\Mstell)$>$10.4).
The models of \cite{Contini2013a} suggest that 68\% of the ICL originates in log(\Mstell)$>$10.5.
This theoretical prediction is not dissimilar from our observational constraints, suggesting that the majority of the ICL originates in such massive galaxies.

\subsection{ICL Origins}

One aspect this study adds to the above narrative, beyond confirming our conclusions from Paper I with a larger sample, is an analysis of how the composition of the ICL changes due to the mass of the system in which it is being formed.
Galaxies might be expected to interact and evolve differently depending on the halo in which they reside.
In more massive clusters ram-pressure stripping, harassment, and minor-mergers are significant mechanisms that affect individual galaxy growth and subsequent evolution \citep{Park2009a, Smith2010a, Wezgowiec2012a}. 
However, in lower-mass groups tidal interaction between galaxies are more efficient at stripping galaxies due to longer interaction times and can therefore significantly impact a galaxy's evolution. 
The ICL records these differences in galaxy interaction rates and types as it collects all of the stripped stars throughout the cluster assembly history. 

To start addressing this concept observationally, we look to how the luminosity, colour, and colour distribution of the ICL behave over the mass range probed in this sample. 
As presented in Table \ref{table:best_fit}, the colour gradient of the ICL remains in the range of $-$0.25$\leq$\nab$<$0 for all systems whether they are measured with an inner fitting radius of 10 or 15 kpc.
However, because of the group colour gradients are not individually robust to the choice of inner fitting radius (See \S\ref{sec:radprofs}), we cannot use the observed colour gradient as a function of total cluster mass as a diagnostic of the dominant formation mechanism. 
Instead we bin our sample into two sub-samples using a dividing mass of 1\tenfourteen\ \Msun\ and simultaneously fit all profiles within a bin.
As discussed in \S\ref{sec:radprofs}, the results of these fits do not indicate that a statistically significant relation exists between colour gradient and total cluster mass.

Given the negative colour gradients, high luminosities, and red colours of the ICL the combined evidence indicate that tidal stripping of massive galaxies (\logmstell$>$10.4) is the dominant channel of ICL formation within 100 kpc (0.08-0.23 \rfive) for groups and clusters spanning 3\tenthirteen$-$9\tenfourteen\ \Msun\ at z$<$0.9.
However, these observations do not isolate whether stripping is predominantly due to the cluster potential or galaxy-galaxy interactions. 
However, we do note that a recent study by \cite{Giallongo2014} has made some basic assumptions about the origin of the ICL and calculated the fraction of cluster light contained in the BCG+ICL as a function of radius. 
Assuming circular orbits, and tidal stripping from the cluster potential alone, they are able to recover the potential shape of cluster CL0024+17 (z\til0.4), as measured with weak lensing.
This correspondence between the weak-lensing potential shape and the distribution of stellar light from the BCG+ICL suggests that stars liberated from their natal galaxies via stripping by the cluster potential (as opposed to galaxy-galaxy interactions) may constitute a significant portion of the total stellar content of the ICL. 
It should be noted that no study has yet to use the distribution of the ICL in a similar manner to test potential formation mechanisms starting with a different baseline assumption (e.g. stripping galaxy-galaxy interaction and/or pre-processing in in-falling groups).

\section{Conclusions}
Using our sample of 23 galaxy groups and clusters ranging in redshift from 0.25$<$z$<$0.89 and \mfive= 3\tenthirteen$-$9\tenfourteen\ \Msun, we constrain the progenitor population and formation mechanism of the ICL with analysis of the surface brightness, colour distribution, total luminosity, and equivalent red sequence galaxy colour.
 
 \begin{enumerate}[(1)]
    \item We rule out major mergers associated with the BCG as the dominant channel of ICL formation beyond 10 kpc. Our \green$-$\red\ colour gradients, fit between either r$>$10 kpc or r$>$15 kpc and 53-110 kpc, are negative for all systems but one at the 3$\sigma$ level or greater.
    Many successive major mergers would eradicate a gradient in the stellar populations of the ICL \citep{La-Barber2012a, Eigenthaler2013a}. 
    Additionally, for violent relaxation after major mergers with BCG to produce the observed level of luminosity of the BCG+ICL at r$<$100 kpc, an order of magnitude too many the expected number of mergers with the BCG would have to occur after z=1 \citep{Lidman2013a}.
    Thus we are left with only tidal stripping or dwarf disruption as potential mechanisms of ICL formation that can produce the observed negative colour gradients in the ICL. 
    
    \item ICL luminosities are too bright for dwarf disruption alone to be the dominant source of intracluster light and leads us to conclude that tidal stripping of more-massive galaxies is the likely dominant formation mechanism of the ICL.
    All CLASH clusters show a total BCG+ICL luminosity within r$<$100 kpc greater than 1.3\tentwelve\ \Lsun, and as high as 3.5\tentwelve\ \Lsun.
    The groups have lower ICL luminosities in the range of 0.7-1.3\tentwelve\ \Lsun.
    To produce the minimum of 6 \Lstar\ of luminosity from 10$<$r$<$100 kpc for the group sample via the disruption of dwarf galaxies would require hundreds of dwarfs. In turn, this would significantly flatten the faint-end slope of the luminosity function for z$<$1, which has not been observed \citep{Mancone2012a,Wylezalek2014}.
    
    \item By matching the colour of the ICL with the colours of cluster galaxy members, we find that the ICL colour within 100 kpc is consistent with red sequence galaxies with M$<$\Mstar+2.5, or L$>$0.1 \Lstar. 
    We have constructed a composite CMD for all CLASH clusters and fit the red-sequence colour-magnitude relation. 
    Under the assumption that a galaxy is completely disrupted and directly added to the ICL, which we note is not physical for massive galaxies, we convert the observed median CLASH BCG+ICL colour profile to the equivalent magnitude red sequence galaxy donors as a function of radius. 
    This offers a conservative lower limit to the mass of the ICL progenitor population. 
    In more realistic scenarios, only the outer regions of galaxies will be stripped during tidal interactions, which are bluer.
    Further, we find that 75\% of the ICL luminosity originates with \Mstar+1.6 and brighter galaxies (\logmstell$>$10.4).
    
    \item We determine that the stellar mass of the inner 100 kpc goes as \logmstell$=$(0.37$\pm$0.05)\logmfive-(6.63$\pm$0.70), which is considerably steeper than the stellar mass of the inner 10 kpc (\logmstell$\propto$(0.17$\pm$0.06)\logmfive) and implies that the inner and outer components are decoupled.
    The stellar content of the BCG+ICL goes up more slowly than the host cluster total mass growth, suggesting that the ICL is more efficiently produced in low-mass, group environments. 
    The similarity of the BCG+ICL absolute colours and colour gradients over our groups and cluster samples indicates that one dominant ICL-producing mechanism operates over our entire mass range.
    Such observations offer a benchmark for future models and simulations to reproduce when investigating the progenitor population and formation mechanism of the ICL in halos ranging from group to cluster masses. 
            
\end{enumerate}

The observed high luminosities, negative colour gradients, and red colour of the ICL point to the following: at z$<$0.9 the ICL in the inner 100 kpc is built up largely via tidally stripping of massive galaxies with \logmstell$>$10.0 (L$>$0.1 \Lstar)
Additionally, we estimate that 75\% of the ICL luminosity in massive clusters like the CLASH sample is consistent in colour with originating in galaxies more massive than log(\Mstell)$>$10.4.
Finally, we find a difference in the total ICL content within 100 kpc in systems at group and cluster masses with more-massive clusters hosting greater total ICL stellar masses.
The wide mass range of this sample (which includes groups and CLASH clusters) makes it one of the first to constrain the progenitor population and formation of the ICL in varying environments for a significant number of groups and clusters, which we use to evaluate the effect of cluster mass on the characteristics of the ICL.
Looking forward, we plan to further constrain the formation history of the ICL by looking to the BCG+ICL content in high redshift (z$>$1) clusters.
Additionally, detailed baryon census measurements of groups and clusters spanning a large mass range offers a way forward in understanding the effect of total cluster mass on the build up and origin of the ICL. 

\section*{Acknowledgments}
We acknowledge support from the National Science Foundation
through grant NSF-1108957. 
Support for Program numbers 12634 and 12575 was provided by NASA through a grant from the Space Telescope Science Institute, which is operated by the Association of Universities for Research in Astronomy, Incorporated, under NASA contract NAS5-26555.
AIZ acknowledges support from NSF grant AST-0908280 and NASA grant ADP-NNX10AD47G.
TC and MD acknowledge partial support of NASA grants HST- GO-12065.01-A. 
The archival and Guest Observer data are based on observations made with the NASA/ESA Hubble Space Telescope which is operated by the Space Telescope Science Institute.

\appendix

\section{Delta-Flats}
\begin{figure*}
    \includegraphics[width=\linewidth]{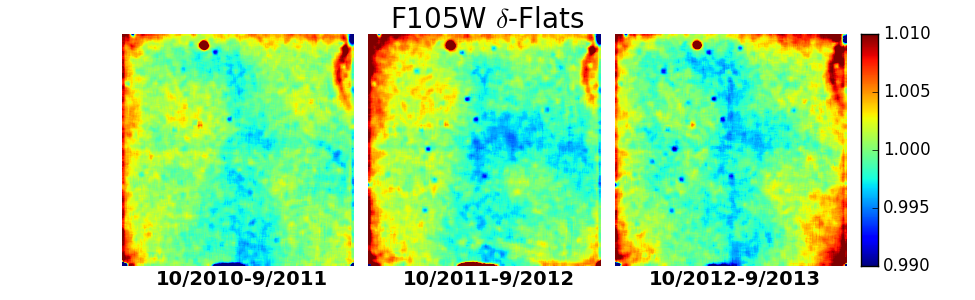}
    \caption{The \blue\ $\delta$-flats produced, marked with the date range of images used to build up the $\delta$-flat. The number of input images in each is detailed in Table \ref{table:flat-epoch}.}
    \label{fig:f105_delta-flat}
\end{figure*}

\begin{figure*}
    \includegraphics[width=\linewidth]{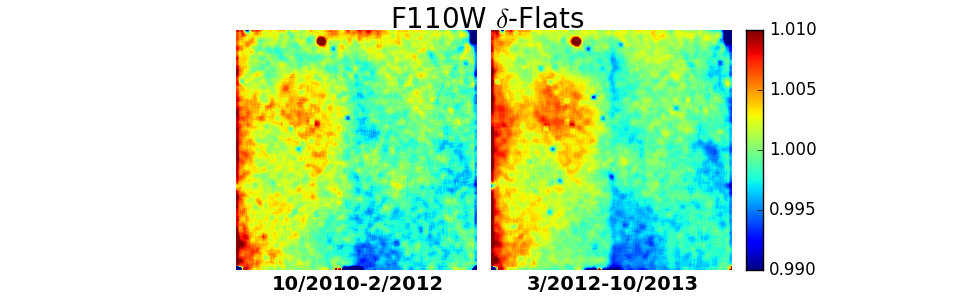}
    \caption{As in Figure \ref{fig:f105_delta-flat}, but for the \green\ $\delta$-flats.}
    \label{fig:f110_delta-flat}
\end{figure*}

\begin{figure*}
    \includegraphics[width=\linewidth]{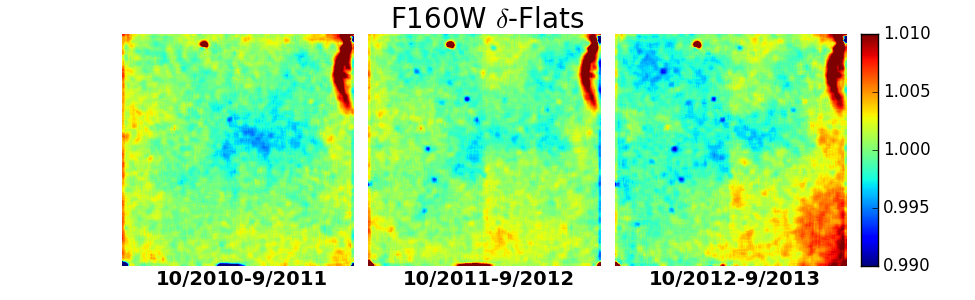}
    \caption{As in Figure \ref{fig:f105_delta-flat}, but for the \red\ $\delta$-flats.}
    \label{fig:f160_delta-flat}
\end{figure*}

To create the $\delta$-flats used in the reduction of \hst\ WFC3/IR imaging, we downloaded all \red, \green, \& \blue\ imaging taken between October 2010--November 2013 from the \hst\ archive.
We then searched through all  observations to isolate only the sparse, extra-galactic fields. 
All Galactic fields, images with high nebulosity, and galaxy clusters were removed and then all astronomical sources were masked in the remaining images. 
Each image was normalised to the median of the unmasked pixels and for a given filter/epoch combination we then median combined the image stack, finally normalizing the combined images once again to the median value.  
Due to the popularity of certain filters for observations, we were only able to create two epochs of $\delta$-flats for \green, whereas for \red\ and \blue\ we had enough images to create three different epochs of $\delta$-flats. 
The number, and date range of the $\delta$-flats which we created are summarised in Table \ref{table:flat-epoch} and images of the delta-flats are shown in Figures \ref{fig:f105_delta-flat}, \ref{fig:f110_delta-flat} and \ref{fig:f160_delta-flat}.

\begin{table}
\centering
\caption{$\delta$-flat Field Date Ranges }
\begin{tabular}{lll}
	Filter & Date Range & \# of Images \\
	\hline
	F160W & 10/2010-9/2011 & 473\\
		  & 10/2011-9/2012 & 632\\
		  & 10/2012-9/2013 & 784\\
	F105W & 10/2010-9/2011 & 194\\
		  & 10/2011-9/2012 & 148\\
		  & 10/2012-9/2013 & 311\\
	F110W & 10/2012-2/2012 & 154\\
		  & 3/2012-10/2013 & 291\\
	\hline
\end{tabular}
\label{table:flat-epoch}
\end{table}

\section{PSFs}
\begin{figure}
    \includegraphics[width=\columnwidth]{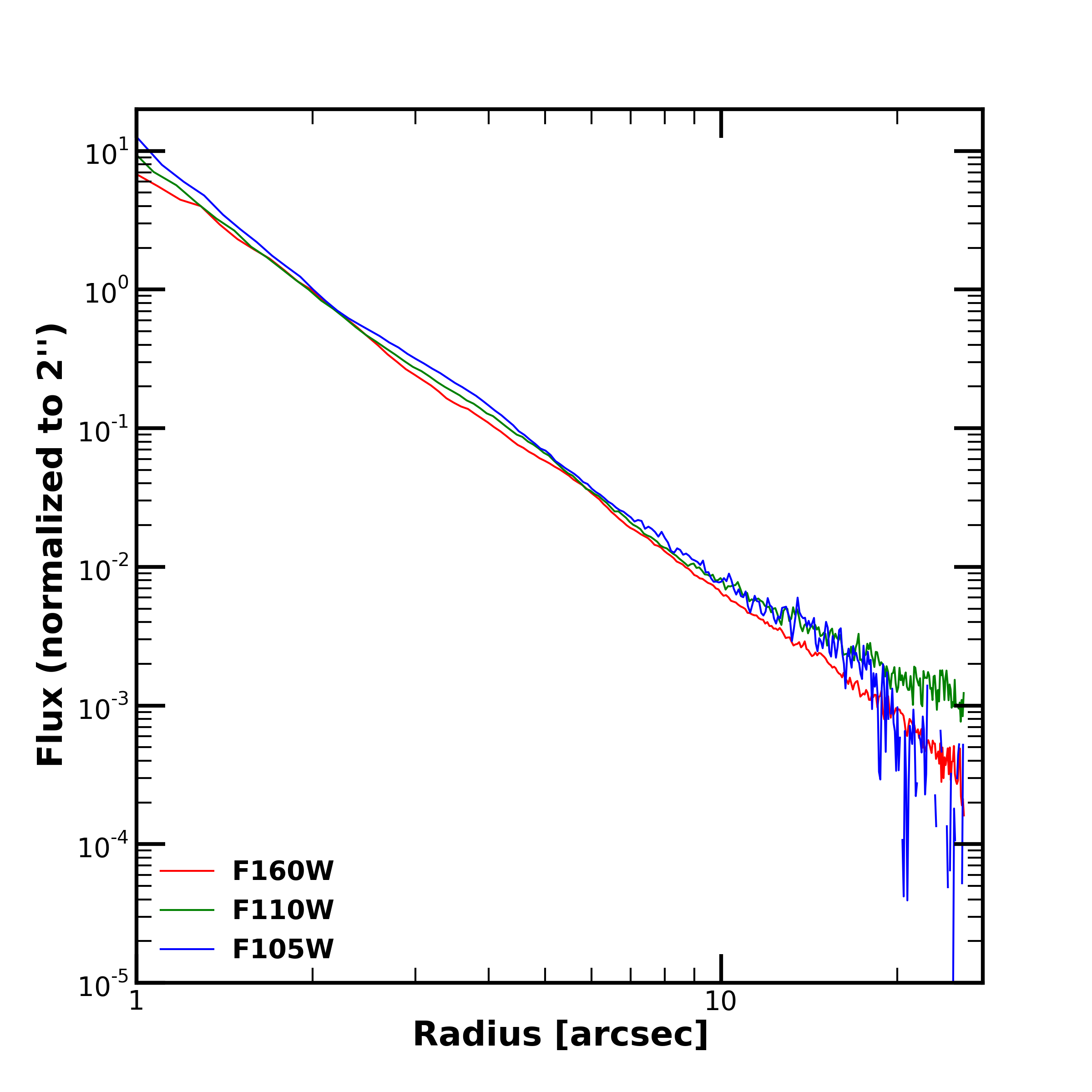}
    \caption{Radial PSFS normalised at 2\arcsec\ out to r=28\arcsec in \blue, \green, and \red\ in blue, green and red lines, respectively. The brightest stars in our science images (\red\til14.5 mag) reach the uncertainty in the background at r$<$25\arcsec, thus these PSFs are sufficient large to ensure than there is not significant contribution to our measurements due to the light in the extended wings of the PSFs.}
    \label{fig:psfs}
\end{figure}

Here we present the \blue, \green, and \red\ PSFs that we create to r$<$28\arcsec.
In Table \ref{table:psf_fields} we list the additional fields used to create the composite PSFs, including the RA, DEC, and program number for each dataset. 
For each filter we use 10-15 fields observed in the same time frame as our science images. 
We present the radially averaged, composite PSFs in each passband, normalised at 2\arcsec, out to r$<$28\arcsec\ in Figure \ref{fig:psfs}.

\begin{table}
\caption{Alternate fields used for PSF creation}
\begin{tabular}{l l l l l}
\hline
Filter & Target Name & Proposal ID & RA & DEC \\
\hline \hline
F160W & ANY & 13767 & 153.7388 & 59.7473 \\
  & ANY & 13767 & 212.4138 & 26.3777 \\
  & ANY & 13767 & 154.3496 & -20.8692 \\
  & SA22A-C30 & 11735 & 334.3303 & 0.2624 \\
  & ANY & 13352 & 255.3878 & 64.1325 \\
  & SSA22A-C6M4 & 11735 & 334.4205 & 0.1908 \\
  & MIPS8495 & 11142 & 258.7041 & 59.8941 \\
  & MIPS549 & 11142 & 259.1203 & 59.4892 \\
  & ANY & 13767 & 154.3496 & -20.8692 \\
  & Q1623-FIELD3 & 11694 & 246.4500 & 26.7427 \\
F110W & ZC415876 & 12578 & 150.0392 & 2.6162 \\
  & ZC412369 & 12578 & 150.4456 & 2.3902 \\
  & SDSS-090740-004160 & 12194 & 136.9168 & -0.7000 \\
  & ZC406690 & 12578 & 149.7464 & 2.0845 \\
  & SDFJ132359.8+272456 & 11149 & 200.9992 & 27.4156 \\
  & D3A15504 & 12578 & 171.0652 & -21.6587 \\
  & GRB-060223 & 11734 & 55.2065 & -17.1301 \\
  & SDFJ132442.5+272423 & 11149 & 201.1771 & 27.4064 \\
  & SDSS-091305-005343 & 12194 & 138.2712 & -0.8952 \\
  & ZC400528 & 12578 & 149.9483 & 1.7386 \\
  & ZC400569 & 12578 & 150.2862 & 1.7412 \\
  & FIELD-142557+354226 & 11153 & 216.4688 & 35.7043 \\
  & ZC409985 & 12578 & 149.8091 & 2.2631 \\
F105W & GRB070802 & 12949 & 36.8995 & -55.5276 \\
  & ANY & 13767 & 258.7510 & 4.9153 \\
  & ANY & 13767 & 212.4138 & 26.3777 \\
  & GOODS-WIDE115-V3T & 12060 & 53.0986 & -27.9002 \\
  & GOODSN-SKIRT000-VDB & 12442 & 189.3848 & 62.2853 \\
  & RARE-FLS-1 & 13718 & 257.0733 & 58.4779 \\
  & F2M1341+3301 & 12942 & 205.2838 & 33.0195 \\
  & ANY & 14096 & 130.7026 & 36.4577 \\
  & ANY & 13767 & 175.5052 & 26.7793 \\
  & ANY & 11584 & 153.5912 & 68.9779 \\
\hline
\hline
\end{tabular}
\end{table}
\label{table:psf_fields}

\section{Cluster Observation Details}
The \hst\ imaging data used for this study comprises \blue, \green, and \red\ images taken over several years from either the CLASH survey or HST-GO \#12575.
In Table \ref{table:sky} we present details on the multi-epoch observations of the CLASH and galaxy group sample, including observation dates, exposure times, background values, and background surface brightness limits for all epochs and all filters of data. 
\onecolumn
\topcaption{Cluster Epoch and Background Values}
\tablefirsthead{\\}
\tablehead{\multicolumn{7}{l}{-- Table \ref{table:sky} continued } \\ \hline }
\begin{xtabular*}{\textwidth}{l@{\extracolsep{\fill}}l l l l l l l}
Cluster & z & Filter & Date & Exposure & Sky & $\delta$ sky \\
        &   &        &      & [s]      & \sbu & \sbu \\
\hline \hline
A611 & 0.29 & F105W & 2012-03-01 & 1306.0 & 21.94 & 28.39 \\
 &  &  & 2012-03-18 & 1509.0 & 21.84 & 27.90 \\
 &  & F110W & 2012-03-02 & 1509.0 & 21.58 & 28.44 \\
 &  &  & 2012-05-17 & 1006.0 & 20.91 & 27.98 \\
 &  & F160W & 2012-03-02 & 1006.0 & 21.73 & 29.00 \\
 &  &  & 2012-05-17 & 1509.0 & 20.86 & 27.98 \\
 &  &  & 2012-03-29 & 1509.0 & 21.55 & 27.63 \\
 &  &  & 2012-01-28 & 1006.0 & 21.55 & 27.90 \\ 
MS2137 & 0.31 & F105W & 2011-09-09 & 1006.0 & 21.72 & 27.82 \\
 &  &  & 2011-10-12 & 703.0 & 21.81 & 28.45 \\
 &  & F110W & 2011-10-20 & 1006.0 & 21.73 & 28.41 \\
 &  &  & 2011-08-21 & 1509.0 & 21.71 & 28.03 \\
 &  & F160W & 2011-08-21 & 1006.0 & 21.66 & 27.75 \\
 &  &  & 2011-09-02 & 1006.0 & 21.68 & 28.19 \\
 &  &  & 2011-10-20 & 1509.0 & 21.47 & 28.30 \\
 &  &  & 2011-11-01 & 1509.0 & 21.21 & 28.01 \\
XMM022045 & 0.33 & F105W & 2012-12-08 & 3012.0 & 22.00 & 29.04 \\
 &  &  & 2012-12-08 & 3012.0 & 22.00 & 29.08 \\
 &  & F160W & 2012-12-08 & 4423.0 & 21.70 & 29.23 \\
RXJ1532 & 0.34 & F105W & 2012-03-16 & 1509.0 & 22.42 & 28.59 \\
 &  &  & 2012-03-03 & 603.0 & 20.85 & 27.38 \\
 &  & F160W & 2012-03-18 & 1509.0 & 22.15 & 27.93 \\
 &  &  & 2012-04-12 & 1509.0 & 22.07 & 27.95 \\
 &  &  & 2012-03-04 & 1006.0 & 22.10 & 26.96 \\
 &  &  & 2012-02-03 & 1006.0 & 21.99 & 28.26 \\
RXJ2248 & 0.35 & F105W & 2012-10-22 & 1306.0 & 22.12 & 28.15 \\
 &  &  & 2012-09-12 & 1509.0 & 22.22 & 27.87 \\
 &  & F110W & 2012-08-30 & 1509.0 & 21.83 & 28.41 \\
 &  &  & 2012-10-04 & 503.0 & 22.02 & 28.08 \\
 &  & F160W & 2012-09-26 & 1006.0 & 21.98 & 27.55 \\
 &  &  & 2012-11-04 & 1509.0 & 21.79 & 27.60 \\
 &  &  & 2012-08-30 & 1006.0 & 22.02 & 27.72 \\
 &  &  & 2012-10-04 & 1509.0 & 21.97 & 27.82 \\
MACS1115 & 0.35 & F105W & 2012-01-31 & 1206.0 & 21.34 & 28.31 \\
 &  &  & 2012-02-23 & 1309.0 & 21.87 & 28.15 \\
 &  & F110W & 2012-02-23 & 503.0 & 21.85 & 27.68 \\
 &  &  & 2011-12-15 & 1309.0 & 21.41 & 27.70 \\
 &  & F160W & 2012-02-24 & 1309.0 & 21.68 & 27.74 \\
 &  &  & 2012-01-31 & 1306.0 & 21.73 & 27.58 \\
 &  &  & 2011-12-15 & 1006.0 & 21.35 & 27.03 \\
 &  &  & 2012-01-07 & 1309.0 & 21.64 & 27.02 \\
MACS1931 & 0.35 & F105W & 2012-05-03 & 1509.0 & 21.68 & 27.19 \\
 &  &  & 2012-06-01 & 1206.0 & 21.81 & 27.87 \\
 &  & F110W & 2012-04-10 & 1509.0 & 21.36 & 27.61 \\
 &  &  & 2012-06-21 & 1006.0 & 20.83 & 27.66 \\
 &  & F160W & 2012-04-10 & 1006.0 & 21.26 & 27.61 \\
 &  &  & 2012-06-21 & 1509.0 & 21.65 & 26.76 \\
 &  &  & 2012-06-25 & 1309.0 & 21.64 & 27.30 \\
 &  &  & 2012-05-03 & 1006.0 & 21.55 & 27.07 \\
SG1120-1 & 0.35 & F105W & 2012-04-28 & 2606.0 & 22.12 & 28.55 \\
 &  &  & 2012-04-28 & 2606.0 & 22.12 & 28.98 \\
 &  & F160W & 2012-04-28 & 5212.0 & 21.84 & 29.10 \\
SG1120-4 & 0.37 & F105W & 2012-06-07 & 3909.0 & 21.85 & 29.19 \\
 &  & F160W & 2012-06-07 & 5212.0 & 21.58 & 28.58 \\
XMM011140 & 0.37 & F105W & 2012-10-19 & 2812.0 & 22.41 & 27.58 \\
 &  &  & 2012-10-19 & 2812.0 & 22.41 & 27.48 \\
 &  & F160W & 2012-10-19 & 4223.0 & 21.95 & 28.37 \\
SG1120-2 & 0.37 & F105W & 2012-06-05 & 3909.0 & 21.88 & 29.17 \\
 &  &  & 2012-06-05 & 3909.0 & 21.88 & 28.65 \\
 &  & F160W & 2012-06-05 & 5212.0 & 21.63 & 28.09 \\
SG1120-3 & 0.37 & F105W & 2012-06-11 & 3909.0 & 21.79 & 28.69 \\
 &  & F160W & 2012-06-11 & 5212.0 & 21.55 & 28.84 \\
RXJ1334 & 0.38 & F105W & 2012-05-02 & 3909.0 & 22.43 & 29.55 \\
 &  &  & 2012-05-02 & 3909.0 & 22.42 & 29.22 \\
 &  & F160W & 2012-05-02 & 5212.0 & 22.05 & 29.34 \\
MACS1720 & 0.39 & F105W & 2012-04-22 & 703.0 & 22.08 & 28.03 \\
 &  &  & 2012-05-09 & 1409.0 & 22.57 & 27.94 \\
 &  & F110W & 2012-04-25 & 1409.0 & 22.52 & 28.51 \\
 &  &  & 2012-06-17 & 503.0 & 22.61 & 28.49 \\
 &  & F160W & 2012-04-25 & 1006.0 & 22.26 & 27.89 \\
 &  &  & 2012-05-05 & 1409.0 & 22.25 & 27.93 \\
 &  &  & 2012-03-26 & 1006.0 & 22.19 & 27.46 \\
 &  &  & 2012-06-17 & 1409.0 & 21.95 & 27.88 \\
MACS0429 & 0.40 & F105W & 2012-12-05 & 1409.0 & 22.10 & 28.51 \\
 &  &  & 2013-01-12 & 1306.0 & 22.16 & 28.51 \\
 &  & F110W & 2012-11-26 & 1409.0 & 22.05 & 28.54 \\
 &  &  & 2012-12-18 & 1006.0 & 22.05 & 28.25 \\
 &  & F160W & 2012-11-26 & 1006.0 & 21.89 & 27.84 \\
 &  &  & 2013-01-27 & 1409.0 & 21.70 & 28.02 \\
 &  &  & 2012-12-11 & 1006.0 & 21.89 & 27.72 \\
 &  &  & 2012-12-18 & 1409.0 & 21.91 & 27.88 \\
RXJ0329 & 0.41 & F105W & 2012-12-05 & 3909.0 & 21.89 & 27.32 \\
 &  & F160W & 2012-12-05 & 5212.0 & 21.77 & 27.01 \\
MACS0416 & 0.42 & F105W & 2012-08-05 & 1509.0 & 21.35 & 28.10 \\
 &  &  & 2012-09-14 & 1306.0 & 21.35 & 27.09 \\
 &  & F160W & 2012-09-27 & 1509.0 & 21.89 & 26.60 \\
 &  &  & 2012-08-31 & 1509.0 & 22.04 & 28.12 \\
 &  &  & 2012-07-24 & 1006.0 & 21.83 & 27.71 \\
 &  &  & 2012-08-05 & 1006.0 & 21.91 & 27.84 \\
MACS0329 & 0.45 & F105W & 2011-10-17 & 703.0 & 21.98 & 28.68 \\
 &  &  & 2011-09-06 & 1509.0 & 21.98 & 28.65 \\
 &  & F110W & 2011-10-16 & 503.0 & 22.00 & 28.69 \\
 &  &  & 2011-08-18 & 1509.0 & 21.80 & 28.56 \\
 &  & F160W & 2011-09-20 & 1006.0 & 21.79 & 27.76 \\
 &  &  & 2011-11-01 & 1509.0 & 21.48 & 28.13 \\
 &  &  & 2011-10-16 & 1509.0 & 21.90 & 28.00 \\
 &  &  & 2011-08-18 & 1006.0 & 21.64 & 28.21 \\
RXJ1347 & 0.45 & F105W & 2011-04-20 & 503.0 & 21.57 & 28.16 \\
 &  &  & 2011-07-12 & 1509.0 & 20.98 & 27.90 \\
 &  & F110W & 2011-07-12 & 1006.0 & 21.51 & 27.74 \\
 &  &  & 2011-04-19 & 1409.0 & 21.53 & 27.92 \\
 &  & F160W & 2011-06-15 & 1306.0 & 21.63 & 27.92 \\
 &  &  & 2011-04-19 & 1006.0 & 21.41 & 28.18 \\
 &  &  & 2011-07-14 & 1509.0 & 21.35 & 28.04 \\
 &  &  & 2011-05-17 & 1509.0 & 21.63 & 28.41 \\
MACS1311 & 0.49 & F105W & 2013-06-10 & 1306.0 & 21.82 & 28.81 \\
 &  &  & 2013-05-18 & 1509.0 & 21.70 & 28.68 \\
 &  & F110W & 2013-07-09 & 1006.0 & 21.45 & 28.40 \\
 &  &  & 2013-04-22 & 1409.0 & 21.62 & 28.60 \\
 &  & F160W & 2013-07-09 & 1509.0 & 21.23 & 28.36 \\
 &  &  & 2013-04-14 & 1006.0 & 21.59 & 28.35 \\
 &  &  & 2013-04-22 & 1006.0 & 21.66 & 28.23 \\
 &  &  & 2013-07-09 & 1509.0 & 21.31 & 28.14 \\
MACS2129 & 0.57 & F105W & 2011-05-16 & 1006.0 & 21.07 & 27.82 \\
 &  &  & 2011-08-03 & 1409.0 & 21.93 & 28.34 \\
 &  & F110W & 2011-07-20 & 503.0 & 21.94 & 28.33 \\
 &  &  & 2011-05-15 & 1409.0 & 21.55 & 28.37 \\
 &  & F160W & 2011-06-25 & 1206.0 & 21.83 & 28.01 \\
 &  &  & 2011-07-20 & 1409.0 & 21.80 & 28.51 \\
 &  &  & 2011-05-15 & 1006.0 & 21.37 & 28.30 \\
 &  &  & 2011-06-03 & 1409.0 & 21.65 & 28.28 \\
CL1226 & 0.89 & F105W & 2013-05-24 & 1609.0 & 22.24 & 28.59 \\
 &  &  & 2013-05-05 & 1206.0 & 22.17 & 28.99 \\
 &  & F110W & 2013-06-22 & 1006.0 & 22.03 & 28.55 \\
 &  &  & 2013-05-09 & 1409.0 & 22.24 & 29.08 \\
 &  & F160W & 2013-06-22 & 1609.0 & 21.78 & 28.64 \\
 &  &  & 2013-04-08 & 1006.0 & 22.02 & 28.75 \\
 &  &  & 2013-05-09 & 1006.0 & 22.04 & 28.78 \\
 &  &  & 2013-05-19 & 1509.0 & 21.89 & 28.73 \\
\hline
\label{table:sky}
\end{xtabular*}
\twocolumn

\section{\blue$-$\red\ colour Correction to \green$-$\red}
\label{sec:color-corr}

To compare the \green$-$\red\ profiles of the CLASH clusters to the \blue$-$\red\ profiles of the galaxy groups, we calculate a redshift-dependent colour transformation for each group and cluster.
We assume that the colour correction increases linearly as the observed \blue$-$\red\ colour becomes redder and that \blue$-$\green=0 when \blue$-$\red=0.
The slope of this relation is then the ratio of \blue$-$\green\ to \blue$-$\red\ for an \Lstar\ galaxy, as produced using \ezgal\ under a Coma normalisation using a BC03 model with a simple stellar population, Chabrier IMF, formation redshift of \zform=3, and solar metallicity.  

\begin{equation}
  \label{eqn:colourcorrection}
  \begin{gathered}
    (\blue-\green) = A \times (\blue-\red) \\
	(\green - \red) = (1-A) \times (\blue-\red)
 \end{gathered}
\end{equation}

The parameter \emph{A} is the colour correction derived from \ezgal, and values used can be found in Table \ref{table:colour-corr}.
Finally, we apply this colour correction, as written in Eqn  \ref{eqn:colourcorrection}, to the \green$-$\red\ colour of each bin in a given cluster's colour profile to produce the final \blue$-$\red\ profile. 
To insure that the choice of model metallicity has no systematic effect on the corrected profiles we use clusters with robust \blue$-$\red\ and \green$-$\red\ profiles to test how  model metallicity affects the colour transformations.
By comparing the actual \green$-$\red\ profiles with converted \green$-$\red\ profiles, which are produced by transforming the \blue$-$\red\ profile with models of varying metallicity.
We find that a colour correction produced with a model of solar or super-solar metallicity is well matched to the observed \green$-$\red\ profile.
Using a solar metallicity model produces transformed colour profiles that have a maximum difference from the observed \green$-$\red\ profiles of 0.04 mag.
Because there is a metallicity gradient in the stellar population of the ICL a single metallicity model does not capture colour transformation perfectly.
Thus we also look at the effects of apply a colour transform assuming a constant metallicity on the measured colour gradients in \green-\red.
We find that the colour gradients measured on the original \green$-$\red\ and transformed profiles are consistent within 3$\sigma$.
We do note that for three of the six clusters with robust \blue$-$\red\ and \green$-$\red\ profiles the transformed profile gradients are shallower than those measured on the observed colour profiles. 

\begin{table}
\centering
\caption{colour Correction Values from \ezgal}
\begin{tabular}{l l}
\hline
Clusters & A \\
\hline
XMM022045 & 0.305 \\
RXJ1532 & 0.289 \\
SG1120-1 & 0.283 \\
SG1120-4 & 0.280 \\
XMM011140 & 0.280 \\
SG1120-2 & 0.280 \\
SG1120-3 & 0.280 \\
RXJ1334 & 0.273 \\
RXJ0329 & 0.253 \\
MACS0416 & 0.247 \\
\hline
\end{tabular}
\label{table:colour-corr}
\end{table}

\section{colour Profiles}
In Table \ref{table:tabular_color}, we present the observed (not e+k corrected) \dlogr=0.15 colour profiles of all systems out to radii where the uncertainty in the measured colour is $<$0.2 \sbu.
These colour profiles are those used in the figures of this paper.
Note, the surface brightness and colour gradient measurements in this paper are derived from the \dlogr=0.05 bin profiles.



\begin{table*}
\caption{colour Profiles}
\centering
\begin{tabular}{cccccccc}
\\
\hline
\multicolumn{2}{c}{A611} & \multicolumn{2}{c}{MS2137} & \multicolumn{2}{c}{XMM022045} & \multicolumn{2}{c}{RXJ1532} \\
log(r[kpc]) & F110W-F160W & log(r[kpc]) & F110W-F160W & log(r[kpc]) & F110W-F160W & log(r[kpc]) & F110W-F160W \\
\hline
0.559 & 0.439$\pm$0.012 & 0.537 & 0.409$\pm$0.016 & 0.540 & 0.375$\pm$0.008 & 0.545 & 0.374$\pm$0.026 \\
0.697 & 0.431$\pm$0.008 & 0.698 & 0.416$\pm$0.004 & 0.689 & 0.390$\pm$0.009 & 0.701 & 0.352$\pm$0.012 \\
0.838 & 0.422$\pm$0.003 & 0.838 & 0.401$\pm$0.007 & 0.837 & 0.403$\pm$0.011 & 0.847 & 0.363$\pm$0.010 \\
0.985 & 0.423$\pm$0.005 & 0.988 & 0.403$\pm$0.007 & 0.987 & 0.401$\pm$0.013 & 0.986 & 0.352$\pm$0.015 \\
1.144 & 0.412$\pm$0.003 & 1.137 & 0.382$\pm$0.002 & 1.132 & 0.386$\pm$0.017 & 1.136 & 0.356$\pm$0.023 \\
1.291 & 0.395$\pm$0.003 & 1.296 & 0.366$\pm$0.006 & 1.287 & 0.378$\pm$0.023 & 1.290 & 0.343$\pm$0.018 \\
1.435 & 0.387$\pm$0.004 & 1.440 & 0.343$\pm$0.007 & 1.436 & 0.367$\pm$0.034 & 1.435 & 0.353$\pm$0.012 \\
1.583 & 0.370$\pm$0.006 & 1.576 & 0.331$\pm$0.017 & 1.589 & 0.327$\pm$0.056 & 1.577 & 0.316$\pm$0.015 \\
1.724 & 0.366$\pm$0.025 & 1.737 & 0.283$\pm$0.045 & 1.727 & 0.306$\pm$0.093 & 1.745 & 0.317$\pm$0.031 \\
1.898 & 0.372$\pm$0.059 & 1.892 & 0.237$\pm$0.089 &   &   & 1.885 & 0.310$\pm$0.068 \\
\hline
\multicolumn{2}{c}{RXJ2248} & \multicolumn{2}{c}{MACS1931} & \multicolumn{2}{c}{MACS1115} & \multicolumn{2}{c}{SG1120-4} \\
log(r[kpc]) & F110W-F160W & log(r[kpc]) & F110W-F160W & log(r[kpc]) & F110W-F160W & log(r[kpc]) & F110W-F160W \\
\hline
0.544 & 0.322$\pm$0.011 & 0.537 & 0.559$\pm$0.008 & 0.535 & 0.400$\pm$0.018 & 0.536 & 0.447$\pm$0.006 \\
0.699 & 0.324$\pm$0.005 & 0.695 & 0.482$\pm$0.023 & 0.693 & 0.416$\pm$0.015 & 0.691 & 0.435$\pm$0.006 \\
0.842 & 0.332$\pm$0.007 & 0.841 & 0.391$\pm$0.013 & 0.850 & 0.419$\pm$0.008 & 0.835 & 0.445$\pm$0.008 \\
0.989 & 0.322$\pm$0.004 & 0.987 & 0.390$\pm$0.007 & 0.990 & 0.414$\pm$0.005 & 0.989 & 0.448$\pm$0.010 \\
1.139 & 0.321$\pm$0.005 & 1.137 & 0.409$\pm$0.003 & 1.135 & 0.410$\pm$0.009 & 1.138 & 0.426$\pm$0.013 \\
1.292 & 0.309$\pm$0.008 & 1.281 & 0.383$\pm$0.015 & 1.287 & 0.391$\pm$0.005 & 1.282 & 0.392$\pm$0.017 \\
1.440 & 0.299$\pm$0.013 & 1.439 & 0.359$\pm$0.034 & 1.435 & 0.396$\pm$0.008 & 1.439 & 0.396$\pm$0.024 \\
1.572 & 0.285$\pm$0.021 & 1.587 & 0.336$\pm$0.054 & 1.581 & 0.364$\pm$0.014 & 1.589 & 0.349$\pm$0.036 \\
1.728 & 0.259$\pm$0.042 & 1.731 & 0.311$\pm$0.091 & 1.723 & 0.319$\pm$0.029 & 1.709 & 0.347$\pm$0.051 \\
1.900 & 0.225$\pm$0.074 &   &   & 1.883 & 0.294$\pm$0.042 & 1.881 & 0.359$\pm$0.091 \\
\hline
\multicolumn{2}{c}{XMM011140} & \multicolumn{2}{c}{SG1120-2} & \multicolumn{2}{c}{SG1120-1} & \multicolumn{2}{c}{SG1120-3} \\
log(r[kpc]) & F110W-F160W & log(r[kpc]) & F110W-F160W & log(r[kpc]) & F110W-F160W & log(r[kpc]) & F110W-F160W \\
\hline
0.533 & 0.435$\pm$0.008 & 0.539 & 0.466$\pm$0.008 & 0.536 & 0.417$\pm$0.009 & 0.538 & 0.455$\pm$0.005 \\
0.691 & 0.435$\pm$0.009 & 0.689 & 0.458$\pm$0.009 & 0.688 & 0.405$\pm$0.011 & 0.688 & 0.454$\pm$0.007 \\
0.840 & 0.432$\pm$0.011 & 0.837 & 0.459$\pm$0.012 & 0.838 & 0.408$\pm$0.015 & 0.834 & 0.435$\pm$0.009 \\
0.987 & 0.429$\pm$0.013 & 0.982 & 0.459$\pm$0.016 & 0.984 & 0.406$\pm$0.019 & 0.989 & 0.439$\pm$0.011 \\
1.137 & 0.412$\pm$0.017 & 1.138 & 0.436$\pm$0.021 & 1.135 & 0.364$\pm$0.025 & 1.138 & 0.422$\pm$0.015 \\
1.288 & 0.386$\pm$0.024 & 1.292 & 0.414$\pm$0.029 & 1.283 & 0.338$\pm$0.033 & 1.285 & 0.395$\pm$0.020 \\
1.435 & 0.371$\pm$0.034 & 1.435 & 0.416$\pm$0.038 & 1.431 & 0.356$\pm$0.044 & 1.436 & 0.399$\pm$0.028 \\
1.575 & 0.336$\pm$0.049 & 1.586 & 0.414$\pm$0.055 & 1.586 & 0.321$\pm$0.066 & 1.581 & 0.370$\pm$0.041 \\
1.725 & 0.290$\pm$0.068 & 1.729 & 0.377$\pm$0.086 &   &   & 1.730 & 0.352$\pm$0.063 \\
1.881 & 0.186$\pm$0.106 &   &   &   &   & 1.866 & 0.339$\pm$0.095 \\
\hline
\multicolumn{2}{c}{RXJ1334} & \multicolumn{2}{c}{MACS1720} & \multicolumn{2}{c}{MACS0429} & \multicolumn{2}{c}{MACS0416} \\
log(r[kpc]) & F110W-F160W & log(r[kpc]) & F110W-F160W & log(r[kpc]) & F110W-F160W & log(r[kpc]) & F110W-F160W \\
\hline
0.538 & 0.459$\pm$0.009 & 0.545 & 0.500$\pm$0.020 & 0.536 & 0.481$\pm$0.008 & 0.542 & 0.494$\pm$0.012 \\
0.688 & 0.424$\pm$0.011 & 0.709 & 0.485$\pm$0.022 & 0.692 & 0.498$\pm$0.008 & 0.689 & 0.492$\pm$0.008 \\
0.835 & 0.435$\pm$0.014 & 0.848 & 0.469$\pm$0.014 & 0.835 & 0.471$\pm$0.007 & 0.838 & 0.491$\pm$0.006 \\
0.989 & 0.429$\pm$0.018 & 0.986 & 0.461$\pm$0.005 & 0.998 & 0.461$\pm$0.006 & 0.990 & 0.483$\pm$0.007 \\
1.134 & 0.408$\pm$0.024 & 1.137 & 0.453$\pm$0.005 & 1.140 & 0.471$\pm$0.006 & 1.137 & 0.480$\pm$0.010 \\
1.288 & 0.375$\pm$0.032 & 1.290 & 0.431$\pm$0.006 & 1.289 & 0.467$\pm$0.006 & 1.286 & 0.452$\pm$0.009 \\
1.442 & 0.368$\pm$0.045 & 1.429 & 0.414$\pm$0.014 & 1.435 & 0.445$\pm$0.008 & 1.428 & 0.429$\pm$0.013 \\
1.581 & 0.334$\pm$0.073 & 1.589 & 0.404$\pm$0.014 & 1.579 & 0.431$\pm$0.019 & 1.590 & 0.390$\pm$0.008 \\
  &   & 1.736 & 0.377$\pm$0.032 & 1.741 & 0.416$\pm$0.030 & 1.735 & 0.354$\pm$0.030 \\
  &   & 1.884 & 0.345$\pm$0.059 & 1.880 & 0.405$\pm$0.018 & 1.878 & 0.364$\pm$0.022 \\
  &   & 2.037 & 0.332$\pm$0.037 & 2.041 & 0.332$\pm$0.031 & 2.050 & 0.352$\pm$0.019 \\
  &   & 2.196 & 0.342$\pm$0.063 & 2.188 & 0.303$\pm$0.064 & 2.187 & 0.344$\pm$0.026 \\
\hline

\end{tabular} 
\label{table:tabular_color}
\end{table*}

\begin{table*}
\caption{Colour Profiles, continued from Table \ref{table:tabular_color}}
\centering
\begin{tabular}{cccccccc}
\\
\hline
\multicolumn{2}{c}{MACS1206} & \multicolumn{2}{c}{MACS0329} & \multicolumn{2}{c}{RXJ1347} & \multicolumn{2}{c}{MACS1311} \\
log(r[kpc]) & F110W-F160W & log(r[kpc]) & F110W-F160W & log(r[kpc]) & F110W-F160W & log(r[kpc]) & F110W-F160W \\
\hline
0.541 & 0.463$\pm$0.018 & 0.541 & 0.464$\pm$0.007 & 0.545 & 0.503$\pm$0.012 & 0.544 & 0.518$\pm$0.012 \\
0.688 & 0.497$\pm$0.013 & 0.686 & 0.438$\pm$0.017 & 0.692 & 0.516$\pm$0.006 & 0.685 & 0.480$\pm$0.013 \\
0.838 & 0.492$\pm$0.012 & 0.838 & 0.451$\pm$0.009 & 0.837 & 0.514$\pm$0.016 & 0.854 & 0.469$\pm$0.015 \\
0.990 & 0.488$\pm$0.009 & 0.992 & 0.437$\pm$0.007 & 0.987 & 0.497$\pm$0.007 & 0.989 & 0.475$\pm$0.010 \\
1.137 & 0.489$\pm$0.015 & 1.132 & 0.447$\pm$0.008 & 1.134 & 0.482$\pm$0.008 & 1.138 & 0.470$\pm$0.013 \\
1.286 & 0.467$\pm$0.009 & 1.293 & 0.428$\pm$0.007 & 1.280 & 0.482$\pm$0.009 & 1.289 & 0.463$\pm$0.008 \\
1.429 & 0.443$\pm$0.015 & 1.430 & 0.433$\pm$0.006 & 1.434 & 0.450$\pm$0.015 & 1.434 & 0.438$\pm$0.006 \\
1.585 & 0.425$\pm$0.015 & 1.599 & 0.442$\pm$0.020 & 1.579 & 0.444$\pm$0.010 & 1.584 & 0.432$\pm$0.016 \\
1.743 & 0.482$\pm$0.058 & 1.734 & 0.408$\pm$0.024 & 1.734 & 0.446$\pm$0.024 & 1.741 & 0.397$\pm$0.020 \\
1.867 & 0.446$\pm$0.112 & 1.873 & 0.401$\pm$0.058 & 1.899 & 0.440$\pm$0.030 & 1.873 & 0.362$\pm$0.036 \\
2.041 & 0.467$\pm$0.105 & 2.039 & 0.400$\pm$0.074 & 2.040 & 0.454$\pm$0.111 & 2.045 & 0.376$\pm$0.054 \\
  &   & 2.187 & 0.420$\pm$0.136 &   &   & 2.193 & 0.266$\pm$0.089 \\
\hline
\multicolumn{2}{c}{MACS1149} & \multicolumn{2}{c}{MACS2129} & \multicolumn{2}{c}{CL1226}\\
log(r[kpc]) & F110W-F160W & log(r[kpc]) & F110W-F160W & log(r[kpc]) & F110W-F160W \\
\cline{1-6} \cline{1-6}
0.537 & 0.509$\pm$0.009 & 0.524 & 0.528$\pm$0.008 & 0.857 & 0.583$\pm$0.017 \\
0.841 & 0.496$\pm$0.015 & 0.836 & 0.526$\pm$0.018 & 1.144 & 0.547$\pm$0.017 \\
0.990 & 0.483$\pm$0.013 & 0.991 & 0.533$\pm$0.024 & 1.292 & 0.576$\pm$0.009 \\
1.140 & 0.478$\pm$0.012 & 1.140 & 0.509$\pm$0.009 & 1.437 & 0.522$\pm$0.018 \\
1.288 & 0.453$\pm$0.014 & 1.284 & 0.494$\pm$0.014 & 1.587 & 0.545$\pm$0.017 \\
1.419 & 0.444$\pm$0.016 & 1.434 & 0.445$\pm$0.022 & 1.732 & 0.520$\pm$0.022 \\
1.593 & 0.443$\pm$0.016 & 1.581 & 0.476$\pm$0.042 & 1.884 & 0.498$\pm$0.045 \\
1.732 & 0.412$\pm$0.039 & 1.732 & 0.437$\pm$0.065 & 2.031 & 0.374$\pm$0.125 \\
\cline{1-6}

\end{tabular}  
\label{table:tabular_color_x2}
\end{table*}

\bibliographystyle{mnras}
\bibliography{PaperInIInIII}

\begin{thebibliography}{}
\makeatletter
\relax
\def\mn@urlcharsother{\let\do\@makeother \do\$\do\&\do\#\do\^\do\_\do\%\do\~}
\def\mn@doi{\begingroup\mn@urlcharsother \@ifnextchar [ {\mn@doi@}
  {\mn@doi@[]}}
\def\mn@doi@[#1]#2{\def\@tempa{#1}\ifx\@tempa\@empty \href
  {http://dx.doi.org/#2} {doi:#2}\else \href {http://dx.doi.org/#2} {#1}\fi
  \endgroup}
\def\mn@eprint#1#2{\mn@eprint@#1:#2::\@nil}
\def\mn@eprint@arXiv#1{\href {http://arxiv.org/abs/#1} {{\tt arXiv:#1}}}
\def\mn@eprint@dblp#1{\href {http://dblp.uni-trier.de/rec/bibtex/#1.xml}
  {dblp:#1}}
\def\mn@eprint@#1:#2:#3:#4\@nil{\def\@tempa {#1}\def\@tempb {#2}\def\@tempc
  {#3}\ifx \@tempc \@empty \let \@tempc \@tempb \let \@tempb \@tempa \fi \ifx
  \@tempb \@empty \def\@tempb {arXiv}\fi \@ifundefined
  {mn@eprint@\@tempb}{\@tempb:\@tempc}{\expandafter \expandafter \csname
  mn@eprint@\@tempb\endcsname \expandafter{\@tempc}}}

\bibitem[\protect\citeauthoryear{{Annunziatella} et~al.,}{{Annunziatella}
  et~al.}{2016}]{Annunziatella2016}
{Annunziatella} M.,  et~al., 2016, \mn@doi [\aap]
  {10.1051/0004-6361/201527399}, \href
  {http://adsabs.harvard.edu/abs/2016A\%26A...585A.160A} {585, A160}

\bibitem[\protect\citeauthoryear{{Ben{\'{\i}}tez}}{{Ben{\'{\i}}tez}}{2000}]{BPZ2000}
{Ben{\'{\i}}tez} N.,  2000, \mn@doi [\apj] {10.1086/308947}, \href
  {http://adsabs.harvard.edu/abs/2000ApJ...536..571B} {536, 571}

\bibitem[\protect\citeauthoryear{{Ben{\'{\i}}tez} et~al.,}{{Ben{\'{\i}}tez}
  et~al.}{2004}]{Benitez2004}
{Ben{\'{\i}}tez} N.,  et~al., 2004, \mn@doi [\apjs] {10.1086/380120}, \href
  {http://adsabs.harvard.edu/abs/2004ApJS..150....1B} {150, 1}

\bibitem[\protect\citeauthoryear{{Bertin} \& {Arnouts}}{{Bertin} \&
  {Arnouts}}{1996}]{SEx}
{Bertin} E.,  {Arnouts} S.,  1996, \aaps, \href
  {http://adsabs.harvard.edu/abs/1996A%26AS..117..393B} {117, 393}

\bibitem[\protect\citeauthoryear{{Brammer}, {Pirzkal}, {McCullough}  \&
  {MacKenty}}{{Brammer} et~al.}{2014}]{HSTHeI}
{Brammer} G.,  {Pirzkal} N.,  {McCullough} P.,   {MacKenty} J.,  2014,
  Technical report, {Time-varying Excess Earth-glow Backgrounds in the WFC3/IR
  Channel}

\bibitem[\protect\citeauthoryear{{Bruzual} \& {Charlot}}{{Bruzual} \&
  {Charlot}}{2003}]{BC03}
{Bruzual} G.,  {Charlot} S.,  2003, \mn@doi [\mnras]
  {10.1046/j.1365-8711.2003.06897.x}, \href
  {http://adsabs.harvard.edu/abs/2003MNRAS.344.1000B} {344, 1000}

\bibitem[\protect\citeauthoryear{{Chabrier}}{{Chabrier}}{2003}]{Chabrier2003}
{Chabrier} G.,  2003, \mn@doi [\pasp] {10.1086/376392}, \href
  {http://adsabs.harvard.edu/abs/2003PASP..115..763C} {115, 763}

\bibitem[\protect\citeauthoryear{{Coe}, {Ben{\'{\i}}tez}, {S{\'a}nchez}, {Jee},
  {Bouwens}  \& {Ford}}{{Coe} et~al.}{2006}]{Coe2006}
{Coe} D.,  {Ben{\'{\i}}tez} N.,  {S{\'a}nchez} S.~F.,  {Jee} M.,  {Bouwens} R.,
    {Ford} H.,  2006, \mn@doi [\aj] {10.1086/505530}, \href
  {http://adsabs.harvard.edu/abs/2006AJ....132..926C} {132, 926}

\bibitem[\protect\citeauthoryear{{Connor} et~al.,}{{Connor}
  et~al.}{2017}]{Connor2017}
{Connor} T.,  et~al., 2017, preprint, \href
  {http://adsabs.harvard.edu/abs/2017arXiv170901925C} {} (\mn@eprint {arXiv}
  {1709.01925})

\bibitem[\protect\citeauthoryear{{Conroy}, {Wechsler}  \& {Kravtsov}}{{Conroy}
  et~al.}{2007}]{Conroy2007}
{Conroy} C.,  {Wechsler} R.~H.,   {Kravtsov} A.~V.,  2007, \mn@doi [\apj]
  {10.1086/521425}, \href {http://adsabs.harvard.edu/abs/2007ApJ...668..826C}
  {668, 826}

\bibitem[\protect\citeauthoryear{{Contini}, {De Lucia}, {Villalobos}  \&
  {Borgani}}{{Contini} et~al.}{2014}]{Contini2013a}
{Contini} E.,  {De Lucia} G.,  {Villalobos} {\'A}.,   {Borgani} S.,  2014,
  \mn@doi [\mnras] {10.1093/mnras/stt2174}, \href
  {http://adsabs.harvard.edu/abs/2014MNRAS.437.3787C} {437, 3787}

\bibitem[\protect\citeauthoryear{{DeMaio}, {Gonzalez}, {Zabludoff}, {Zaritsky}
  \& {Brada{\v c}}}{{DeMaio} et~al.}{2015}]{DeMaio2015}
{DeMaio} T.,  {Gonzalez} A.~H.,  {Zabludoff} A.,  {Zaritsky} D.,   {Brada{\v
  c}} M.,  2015, \mn@doi [\mnras] {10.1093/mnras/stv033}, \href
  {http://adsabs.harvard.edu/abs/2015MNRAS.448.1162D} {448, 1162}

\bibitem[\protect\citeauthoryear{{Di Matteo}, {Pipino}, {Lehnert}, {Combes}  \&
  {Semelin}}{{Di Matteo} et~al.}{2009}]{Di-Matteo2009a}
{Di Matteo} P.,  {Pipino} A.,  {Lehnert} M.~D.,  {Combes} F.,   {Semelin} B.,
  2009, \mn@doi [\aap] {10.1051/0004-6361/200911715}, \href
  {http://adsabs.harvard.edu/abs/2009A%26A...499..427D} {499, 427}

\bibitem[\protect\citeauthoryear{{Dressler}}{{Dressler}}{1980}]{Dressler1980}
{Dressler} A.,  1980, \mn@doi [\apj] {10.1086/157753}, \href
  {http://adsabs.harvard.edu/abs/1980ApJ...236..351D} {236, 351}

\bibitem[\protect\citeauthoryear{{Eigenthaler} \& {Zeilinger}}{{Eigenthaler} \&
  {Zeilinger}}{2013}]{Eigenthaler2013a}
{Eigenthaler} P.,  {Zeilinger} W.~W.,  2013, \mn@doi [\aap]
  {10.1051/0004-6361/201321078}, \href
  {http://adsabs.harvard.edu/abs/2013A%26A...553A..99E} {553, A99}

\bibitem[\protect\citeauthoryear{{Giallongo} et~al.,}{{Giallongo}
  et~al.}{2014}]{Giallongo2014}
{Giallongo} E.,  et~al., 2014, \mn@doi [\apj] {10.1088/0004-637X/781/1/24},
  \href {http://adsabs.harvard.edu/abs/2014ApJ...781...24G} {781, 24}

\bibitem[\protect\citeauthoryear{{Gonzalez}, {Tran}, {Conbere}  \&
  {Zaritsky}}{{Gonzalez} et~al.}{2005}]{Gonzalez2005b}
{Gonzalez} A.~H.,  {Tran} K.-V.~H.,  {Conbere} M.~N.,   {Zaritsky} D.,  2005,
  \mn@doi [\apjl] {10.1086/430518}, \href
  {http://adsabs.harvard.edu/abs/2005ApJ...624L..73G} {624, L73}

\bibitem[\protect\citeauthoryear{{Gonzalez}, {Sivanandam}, {Zabludoff}  \&
  {Zaritsky}}{{Gonzalez} et~al.}{2013}]{Gonzalez2013a}
{Gonzalez} A.~H.,  {Sivanandam} S.,  {Zabludoff} A.~I.,   {Zaritsky} D.,  2013,
  \apj, \href {http://adsabs.harvard.edu/abs/2013ApJ...778...14G} {778, 14}

\bibitem[\protect\citeauthoryear{{Hinshaw} et~al.,}{{Hinshaw}
  et~al.}{2013}]{WMAP9}
{Hinshaw} G.,  et~al., 2013, \mn@doi [\apjs] {10.1088/0067-0049/208/2/19},
  \href {http://adsabs.harvard.edu/abs/2013ApJS..208...19H} {208, 19}

\bibitem[\protect\citeauthoryear{{Jee}}{{Jee}}{2010}]{Jee2010}
{Jee} M.~J.,  2010, \mn@doi [\apj] {10.1088/0004-637X/717/1/420}, \href
  {http://adsabs.harvard.edu/abs/2010ApJ...717..420J} {717, 420}

\bibitem[\protect\citeauthoryear{{Jeltema}, {Mulchaey}, {Lubin}, {Rosati}  \&
  {B{\"o}hringer}}{{Jeltema} et~al.}{2006}]{Jeltema2006}
{Jeltema} T.~E.,  {Mulchaey} J.~S.,  {Lubin} L.~M.,  {Rosati} P.,
  {B{\"o}hringer} H.,  2006, \mn@doi [\apj] {10.1086/506372}, \href
  {http://adsabs.harvard.edu/abs/2006ApJ...649..649J} {649, 649}

\bibitem[\protect\citeauthoryear{{Kobayashi}}{{Kobayashi}}{2004}]{Kobayashi2004a}
{Kobayashi} C.,  2004, \mn@doi [\mnras] {10.1111/j.1365-2966.2004.07258.x},
  \href {http://adsabs.harvard.edu/abs/2004MNRAS.347..740K} {347, 740}

\bibitem[\protect\citeauthoryear{{Krick} \& {Bernstein}}{{Krick} \&
  {Bernstein}}{2007}]{KrickII}
{Krick} J.~E.,  {Bernstein} R.~A.,  2007, \mn@doi [\aj] {10.1086/518787}, \href
  {http://adsabs.harvard.edu/abs/2007AJ....134..466K} {134, 466}

\bibitem[\protect\citeauthoryear{{Krick}, {Bernstein}  \& {Pimbblet}}{{Krick}
  et~al.}{2006}]{KrickI}
{Krick} J.~E.,  {Bernstein} R.~A.,   {Pimbblet} K.~A.,  2006, \mn@doi [\aj]
  {10.1086/498269}, \href {http://adsabs.harvard.edu/abs/2006AJ....131..168K}
  {131, 168}

\bibitem[\protect\citeauthoryear{{Krist}, {Hook}  \& {Stoehr}}{{Krist}
  et~al.}{2011}]{TinyTim}
{Krist} J.~E.,  {Hook} R.~N.,   {Stoehr} F.,  2011, in Optical Modeling and
  Performance Predictions V. p. 81270J, \mn@doi{10.1117/12.892762}

\bibitem[\protect\citeauthoryear{{Kuntschner} et~al.,}{{Kuntschner}
  et~al.}{2010}]{Kuntschner2010}
{Kuntschner} H.,  et~al., 2010, \mn@doi [\mnras]
  {10.1111/j.1365-2966.2010.17161.x}, \href
  {http://adsabs.harvard.edu/abs/2010MNRAS.408...97K} {408, 97}

\bibitem[\protect\citeauthoryear{{La Barbera}, {De Carvalho}, {De La Rosa},
  {Gal}, {Swindle}  \& {Lopes}}{{La Barbera} et~al.}{2010}]{La-Barber2010a}
{La Barbera} F.,  {De Carvalho} R.~R.,  {De La Rosa} I.~G.,  {Gal} R.~R.,
  {Swindle} R.,   {Lopes} P.~A.~A.,  2010, \mn@doi [\aj]
  {10.1088/0004-6256/140/5/1528}, \href
  {http://adsabs.harvard.edu/abs/2010AJ....140.1528L} {140, 1528}

\bibitem[\protect\citeauthoryear{{La Barbera}, {Ferreras}, {de Carvalho},
  {Bruzual}, {Charlot}, {Pasquali}  \& {Merlin}}{{La Barbera}
  et~al.}{2012}]{La-Barber2012a}
{La Barbera} F.,  {Ferreras} I.,  {de Carvalho} R.~R.,  {Bruzual} G.,
  {Charlot} S.,  {Pasquali} A.,   {Merlin} E.,  2012, \mn@doi [\mnras]
  {10.1111/j.1365-2966.2012.21848.x}, \href
  {http://adsabs.harvard.edu/abs/2012MNRAS.426.2300L} {426, 2300}

\bibitem[\protect\citeauthoryear{{Lagan{\'a}}, {Martinet}, {Durret}, {Lima
  Neto}, {Maughan}  \& {Zhang}}{{Lagan{\'a}} et~al.}{2013}]{Lagana2013}
{Lagan{\'a}} T.~F.,  {Martinet} N.,  {Durret} F.,  {Lima Neto} G.~B.,
  {Maughan} B.,   {Zhang} Y.-Y.,  2013, \mn@doi [\aap]
  {10.1051/0004-6361/201220423}, \href
  {http://adsabs.harvard.edu/abs/2013A%26A...555A..66L} {555, A66}

\bibitem[\protect\citeauthoryear{{Lidman} et~al.,}{{Lidman}
  et~al.}{2013}]{Lidman2013a}
{Lidman} C.,  et~al., 2013, \mn@doi [\mnras] {10.1093/mnras/stt777}, \href
  {http://adsabs.harvard.edu/abs/2013MNRAS.433..825L} {433, 825}

\bibitem[\protect\citeauthoryear{{Limousin} et~al.,}{{Limousin}
  et~al.}{2012}]{Limousin2012}
{Limousin} M.,  et~al., 2012, \mn@doi [\aap] {10.1051/0004-6361/201117921},
  \href {http://adsabs.harvard.edu/abs/2012A%26A...544A..71L} {544, A71}

\bibitem[\protect\citeauthoryear{{Lin}, {Mohr}  \& {Stanford}}{{Lin}
  et~al.}{2004}]{Lin2004a}
{Lin} Y.-T.,  {Mohr} J.~J.,   {Stanford} S.~A.,  2004, \mn@doi [\apj]
  {10.1086/421714}, \href {http://adsabs.harvard.edu/abs/2004ApJ...610..745L}
  {610, 745}

\bibitem[\protect\citeauthoryear{{Mahdavi}, {Hoekstra}, {Bildfell}, {Jeltema}
  \& {Henry}}{{Mahdavi} et~al.}{2013}]{Mahdavi2013}
{Mahdavi} A.,  {Hoekstra} H.,  {Bildfell} C.,  {Jeltema} T.,   {Henry} J.~P.,
  2013, \mn@doi [\apj] {10.1088/0004-637X/763/1/18}, \href
  {http://adsabs.harvard.edu/abs/2013ApJ...763...18B} {767, 116}

\bibitem[\protect\citeauthoryear{{Mancone} \& {Gonzalez}}{{Mancone} \&
  {Gonzalez}}{2012}]{ezgal}
{Mancone} C.,  {Gonzalez} A.,  2012, preprint, \href
  {http://adsabs.harvard.edu/abs/2012arXiv1205.0009M} {} (\mn@eprint {arXiv}
  {1205.0009})

\bibitem[\protect\citeauthoryear{{Mancone} et~al.,}{{Mancone}
  et~al.}{2012}]{Mancone2012a}
{Mancone} C.~L.,  et~al., 2012, \mn@doi [\apj] {10.1088/0004-637X/761/2/141},
  \href {http://adsabs.harvard.edu/abs/2012ApJ...761..141M} {761, 141}

\bibitem[\protect\citeauthoryear{{Mehrtens} et~al.,}{{Mehrtens}
  et~al.}{2012}]{Mehrtens2012}
{Mehrtens} N.,  et~al., 2012, \mn@doi [\mnras]
  {10.1111/j.1365-2966.2012.20931.x}, \href
  {http://adsabs.harvard.edu/abs/2012MNRAS.423.1024M} {423, 1024}

\bibitem[\protect\citeauthoryear{{Melnick}, {Giraud}, {Toledo}, {Selman}  \&
  {Quintana}}{{Melnick} et~al.}{2012}]{Melnick2012}
{Melnick} J.,  {Giraud} E.,  {Toledo} I.,  {Selman} F.~J.,   {Quintana} H.,
  2012, preprint, \href {http://adsabs.harvard.edu/abs/2012arXiv1207.6394M} {}
  (\mn@eprint {arXiv} {1207.6394})

\bibitem[\protect\citeauthoryear{{Montes} \& {Trujillo}}{{Montes} \&
  {Trujillo}}{2014}]{Montes2014}
{Montes} M.,  {Trujillo} I.,  2014, \mn@doi [\apj]
  {10.1088/0004-637X/794/2/137}, \href
  {http://adsabs.harvard.edu/abs/2014ApJ...794..137M} {794, 137}

\bibitem[\protect\citeauthoryear{{Morishita}, {Abramson}, {Treu}, {Schmidt},
  {Vulcani}  \& {Wang}}{{Morishita} et~al.}{2016}]{Morishita2016}
{Morishita} T.,  {Abramson} L.~E.,  {Treu} T.,  {Schmidt} K.~B.,  {Vulcani} B.,
    {Wang} X.,  2016, preprint, \href
  {http://adsabs.harvard.edu/abs/2016arXiv161008503M} {} (\mn@eprint {arXiv}
  {1610.08503})

\bibitem[\protect\citeauthoryear{{Moustakas} et~al.,}{{Moustakas}
  et~al.}{2013}]{Moustakas2013}
{Moustakas} J.,  et~al., 2013, \mn@doi [\apj] {10.1088/0004-637X/767/1/50},
  \href {http://adsabs.harvard.edu/abs/2013ApJ...767...50M} {767, 50}

\bibitem[\protect\citeauthoryear{{Mulchaey}, {Lubin}, {Fassnacht}, {Rosati}  \&
  {Jeltema}}{{Mulchaey} et~al.}{2006}]{Mulchaey2006a}
{Mulchaey} J.~S.,  {Lubin} L.~M.,  {Fassnacht} C.,  {Rosati} P.,   {Jeltema}
  T.~E.,  2006, \apj, \href {http://adsabs.harvard.edu/abs/2006ApJ...646..133M}
  {646, 133}

\bibitem[\protect\citeauthoryear{{Murante}, {Giovalli}, {Gerhard}, {Arnaboldi},
  {Borgani}  \& {Dolag}}{{Murante} et~al.}{2007}]{Murante2007}
{Murante} G.,  {Giovalli} M.,  {Gerhard} O.,  {Arnaboldi} M.,  {Borgani} S.,
  {Dolag} K.,  2007, \mn@doi [\mnras] {10.1111/j.1365-2966.2007.11568.x}, \href
  {http://adsabs.harvard.edu/abs/2007MNRAS.377....2M} {377, 2}

\bibitem[\protect\citeauthoryear{{Muzzin}, {Yee}, {Hall}, {Ellingson}  \&
  {Lin}}{{Muzzin} et~al.}{2007}]{Muzzin2007}
{Muzzin} A.,  {Yee} H.~K.~C.,  {Hall} P.~B.,  {Ellingson} E.,   {Lin} H.,
  2007, \mn@doi [\apj] {10.1086/511669}, \href
  {http://adsabs.harvard.edu/abs/2007ApJ...659.1106M} {659, 1106}

\bibitem[\protect\citeauthoryear{{Park} \& {Hwang}}{{Park} \&
  {Hwang}}{2009}]{Park2009a}
{Park} C.,  {Hwang} H.~S.,  2009, \mn@doi [\apj]
  {10.1088/0004-637X/699/2/1595}, \href
  {http://adsabs.harvard.edu/abs/2009ApJ...699.1595P} {699, 1595}

\bibitem[\protect\citeauthoryear{{Pirzkal}, {Mack}, {Dahlen}  \&
  {Sabbi}}{{Pirzkal} et~al.}{2011}]{Pirzkal2011}
{Pirzkal} N.,  {Mack} J.,  {Dahlen} T.,   {Sabbi} E.,  2011, Technical report,
  {Sky Flats: Generating Improved WFC3 IR Flat-fields}

\bibitem[\protect\citeauthoryear{{Postman} et~al.,}{{Postman}
  et~al.}{2012a}]{Postman2012a}
{Postman} M.,  et~al., 2012a, \mn@doi [\apjs] {10.1088/0067-0049/199/2/25},
  \href {http://adsabs.harvard.edu/abs/2012ApJS..199...25P} {199, 25}

\bibitem[\protect\citeauthoryear{{Postman} et~al.,}{{Postman}
  et~al.}{2012b}]{Postman2012}
{Postman} M.,  et~al., 2012b, \mn@doi [\apj] {10.1088/0004-637X/756/2/159},
  \href {http://adsabs.harvard.edu/abs/2012ApJ...756..159P} {756, 159}

\bibitem[\protect\citeauthoryear{{Rudick}, {Mihos}  \& {McBride}}{{Rudick}
  et~al.}{2006}]{Rudick2006}
{Rudick} C.~S.,  {Mihos} J.~C.,   {McBride} C.,  2006, \mn@doi [\apj]
  {10.1086/506176}, \href {http://adsabs.harvard.edu/abs/2006ApJ...648..936R}
  {648, 936}

\bibitem[\protect\citeauthoryear{{Rudick}, {Mihos}, {Harding}, {Feldmeier},
  {Janowiecki}  \& {Morrison}}{{Rudick} et~al.}{2010}]{Rudick2010}
{Rudick} C.~S.,  {Mihos} J.~C.,  {Harding} P.,  {Feldmeier} J.~J.,
  {Janowiecki} S.,   {Morrison} H.~L.,  2010, \mn@doi [\apj]
  {10.1088/0004-637X/720/1/569}, \href
  {http://adsabs.harvard.edu/abs/2010ApJ...720..569R} {720, 569}

\bibitem[\protect\citeauthoryear{{Skillman}, {Kennicutt}, {Shields}  \&
  {Zaritsky}}{{Skillman} et~al.}{1996}]{Skillman1996a}
{Skillman} E.~D.,  {Kennicutt} Jr. R.~C.,  {Shields} G.~A.,   {Zaritsky} D.,
  1996, \mn@doi [\apj] {10.1086/177138}, \href
  {http://adsabs.harvard.edu/abs/1996ApJ...462..147S} {462, 147}

\bibitem[\protect\citeauthoryear{{Smith}, {Davies}  \& {Nelson}}{{Smith}
  et~al.}{2010}]{Smith2010a}
{Smith} R.,  {Davies} J.~I.,   {Nelson} A.~H.,  2010, \mn@doi [\mnras]
  {10.1111/j.1365-2966.2010.16545.x}, \href
  {http://adsabs.harvard.edu/abs/2010MNRAS.405.1723S} {405, 1723}

\bibitem[\protect\citeauthoryear{{Strazzullo} et~al.,}{{Strazzullo}
  et~al.}{2010}]{Strazzullo2010}
{Strazzullo} V.,  et~al., 2010, \mn@doi [\aap] {10.1051/0004-6361/201015251},
  \href {http://adsabs.harvard.edu/abs/2010A%26A...524A..17S} {524, A17}

\bibitem[\protect\citeauthoryear{{Vikhlinin} et~al.,}{{Vikhlinin}
  et~al.}{2009}]{Vikhlinin2009}
{Vikhlinin} A.,  et~al., 2009, \mn@doi [\apj] {10.1088/0004-637X/692/2/1033},
  \href {http://adsabs.harvard.edu/abs/2009ApJ...692.1033V} {692, 1033}

\bibitem[\protect\citeauthoryear{{We{\.z}gowiec}, {Bomans}, {Ehle},
  {Chy{\.z}y}, {Urbanik}, {Braine}  \& {Soida}}{{We{\.z}gowiec}
  et~al.}{2012}]{Wezgowiec2012a}
{We{\.z}gowiec} M.,  {Bomans} D.~J.,  {Ehle} M.,  {Chy{\.z}y} K.~T.,  {Urbanik}
  M.,  {Braine} J.,   {Soida} M.,  2012, \mn@doi [\aap]
  {10.1051/0004-6361/201117652}, \href
  {http://adsabs.harvard.edu/abs/2012A%26A...544A..99W} {544, A99}

\bibitem[\protect\citeauthoryear{{Wylezalek} et~al.,}{{Wylezalek}
  et~al.}{2014}]{Wylezalek2014}
{Wylezalek} D.,  et~al., 2014, \mn@doi [\apj] {10.1088/0004-637X/786/1/17},
  \href {http://adsabs.harvard.edu/abs/2014ApJ...786...17W} {786, 17}

\bibitem[\protect\citeauthoryear{{Zaritsky}, {Kennicutt}  \&
  {Huchra}}{{Zaritsky} et~al.}{1994}]{Zaritsky1994a}
{Zaritsky} D.,  {Kennicutt} Jr. R.~C.,   {Huchra} J.~P.,  1994, \mn@doi [\apj]
  {10.1086/173544}, \href {http://adsabs.harvard.edu/abs/1994ApJ...420...87Z}
  {420, 87}

\bibitem[\protect\citeauthoryear{{Zhang}, {Finoguenov}, {B{\"o}hringer},
  {Kneib}, {Smith}, {Czoske}  \& {Soucail}}{{Zhang} et~al.}{2007}]{Zhang2007}
{Zhang} Y.-Y.,  {Finoguenov} A.,  {B{\"o}hringer} H.,  {Kneib} J.-P.,  {Smith}
  G.~P.,  {Czoske} O.,   {Soucail} G.,  2007, \mn@doi [\aap]
  {10.1051/0004-6361:20066567}, \href
  {http://adsabs.harvard.edu/abs/2007A%26A...467..437Z} {467, 437}

\bibitem[\protect\citeauthoryear{{Ziparo} et~al.,}{{Ziparo}
  et~al.}{2013}]{Ziparo2013}
{Ziparo} F.,  et~al., 2013, \mn@doi [\mnras] {10.1093/mnras/stt1222}, \href
  {http://adsabs.harvard.edu/abs/2013MNRAS.434.3089Z} {434, 3089}

\makeatother
\end{thebibliography}

\end{document}